\documentclass[a4paper]{article}
\usepackage{graphicx}
\usepackage[margin=1in]{geometry}
\usepackage{xcolor}
\usepackage{floatrow}
\usepackage{multirow}
\usepackage{comment}
\usepackage{ulem} 

\usepackage{hyperref}
\hypersetup{
    colorlinks=true,
    linkcolor=black, 
    filecolor=black,      
    urlcolor=black,
    citecolor = black
    }

\usepackage{floatrow}
\usepackage{ifthen}
\newboolean{pdflatex}
\setboolean{pdflatex}{false} 

\newboolean{uprightparticles}
\setboolean{uprightparticles}{false} 

\newboolean{inbibliography}
\setboolean{inbibliography}{false} 

\newboolean{articletitles}
\setboolean{articletitles}{true} 

\usepackage{subcaption}
\usepackage{placeins}

\usepackage[mathlines,displaymath]{lineno}

\usepackage{xspace} 
\usepackage{upgreek}

\def\MagUp {\mbox{\em Mag\kern -0.05em Up}\xspace}

\ifthenelse{\boolean{uprightparticles}}%
{

 \def\PDelta      {\ensuremath{\Delta}\xspace}                 
 \def\PXi         {\ensuremath{\Xi}\xspace}                 
 \def\PLambda     {\ensuremath{\Lambda}\xspace}                 
 \def\PSigma      {\ensuremath{\Sigma}\xspace}                 
 \def\POmega      {\ensuremath{\Omega}\xspace}                 
 \def\PUpsilon    {\ensuremath{\Upsilon}\xspace}

 \def\PB      {\ensuremath{\mathrm{B}}\xspace}                 
                  
 \def\PD      {\ensuremath{\mathrm{D}}\xspace}

 \def\PK      {\ensuremath{\mathrm{K}}\xspace}

 \def\Pi      {\ensuremath{\mathrm{i}}\xspace}

 \def\Ps      {\ensuremath{\mathrm{s}}\xspace}

 \def\thebaroffset{0.0em}
}
{

 \mathchardef\PDelta="7101
 \mathchardef\PXi="7104
 \mathchardef\PLambda="7103
 \mathchardef\PSigma="7106
 \mathchardef\POmega="710A
 \mathchardef\PUpsilon="7107
                  
 \def\PB      {\ensuremath{B}\xspace}                 
                  
 \def\PD      {\ensuremath{D}\xspace}

 \def\PK      {\ensuremath{K}\xspace}

 \def\Pi      {\ensuremath{i}\xspace}

 \def\Ps      {\ensuremath{s}\xspace}

 \def\thebaroffset{0.18em}
}
\newcommand{\offsetoverline}[2][\thebaroffset]{\kern #1\overline{\kern -#1 #2}}%

\makeatletter
\ifcase \@ptsize \relax
  \newcommand{\miniscule}{\@setfontsize\miniscule{4}{5}}
\or
  \newcommand{\miniscule}{\@setfontsize\miniscule{5}{6}}
\or
  \newcommand{\miniscule}{\@setfontsize\miniscule{5}{6}}
\fi
\makeatother

\DeclareRobustCommand{\optbar}[1]{\shortstack{{\miniscule (\rule[.5ex]{1.25em}{.18mm})}
  \\ [-.7ex] $#1$}}

\def\squark    {{\ensuremath{\Ps}}\xspace}

\def\KorKbar {\kern \thebaroffset\optbar{\kern -\thebaroffset \PK}{}\xspace}

\def\DorDbar {\kern \thebaroffset\optbar{\kern -\thebaroffset \PD}\xspace}

\def\B       {{\ensuremath{\PB}}\xspace}

\def\BorBbar {\kern \thebaroffset\optbar{\kern -\thebaroffset \PB}\xspace}

\def\Bd      {{\ensuremath{\B^0}}\xspace}

\def\BdorBdbar {\kern \thebaroffset\optbar{\kern -\thebaroffset \Bd}\xspace}

\def\Bs      {{\ensuremath{\B^0_\squark}}\xspace}

\def\BsorBsbar {\kern \thebaroffset\optbar{\kern -\thebaroffset \Bs}\xspace}

\def\Y#1S{\ensuremath{\PUpsilon{(#1S)}}\xspace}

\def\LorLbar     {\kern \thebaroffset\optbar{\kern -\thebaroffset \PLambda}\xspace}

\def\AT#1     {\ensuremath{A_{\mathrm{T}}^{#1}}\xspace}           

\def\C#1      {\ensuremath{\mathcal{C}_{#1}}\xspace}                       
\def\Cp#1     {\ensuremath{\mathcal{C}_{#1}^{'}}\xspace}                    
\def\Ceff#1   {\ensuremath{\mathcal{C}_{#1}^{\mathrm{(eff)}}}\xspace}        
\def\Cpeff#1  {\ensuremath{\mathcal{C}_{#1}^{'\mathrm{(eff)}}}\xspace}       
\def\Ope#1    {\ensuremath{\mathcal{O}_{#1}}\xspace}                       
\def\Opep#1   {\ensuremath{\mathcal{O}_{#1}^{'}}\xspace}                    

\newcommand{\aunit}[1]{\ensuremath{\text{\,#1}}}       

\newcommand{\tev}{\aunit{Te\kern -0.1em V}\xspace}
\newcommand{\gev}{\aunit{Ge\kern -0.1em V}\xspace}
\newcommand{\mev}{\aunit{Me\kern -0.1em V}\xspace}
\newcommand{\kev}{\aunit{ke\kern -0.1em V}\xspace}
\newcommand{\ev}{\aunit{e\kern -0.1em V}\xspace}
\newcommand{\mevc}{\ensuremath{\aunit{Me\kern -0.1em V\!/}c}\xspace}
\newcommand{\gevc}{\ensuremath{\aunit{Ge\kern -0.1em V\!/}c}\xspace}
\newcommand{\mevcc}{\ensuremath{\aunit{Me\kern -0.1em V\!/}c^2}\xspace}
\newcommand{\gevcc}{\ensuremath{\aunit{Ge\kern -0.1em V\!/}c^2}\xspace}

\def\m    {\aunit{m}\xspace}

\def\mm   {\aunit{mm}\xspace}
\def\mma  {\ensuremath{\aunit{mm}^2}\xspace}

\def\nm   {\aunit{nm}\xspace}

\def\ns   {\ensuremath{\aunit{ns}}\xspace}
\def\ps   {\ensuremath{\aunit{ps}}\xspace}

\def\gsim{{~\raise.15em\hbox{$>$}\kern-.85em
          \lower.35em\hbox{$\sim$}~}\xspace}
\def\lsim{{~\raise.15em\hbox{$<$}\kern-.85em
          \lower.35em\hbox{$\sim$}~}\xspace}

\def\tell1  {TELL1\xspace}
\def\ukl1   {UKL1\xspace}

\def\mmv {\ensuremath{\aunit{mm}^3}\xspace}

\makeatletter
\renewcommand*{\@fnsymbol}[1]{\ensuremath{\ifcase#1\or 1 \or 2 \or 3\or * \else\@ctrerr\fi}}
\makeatother

\title{Performance of a prototype TORCH time-of-flight detector} 

\author{ 
S.~Bhasin$^{a,b}$,
T.~Blake$^c$, 
N.~H.~Brook$^{b,}$\thanks{Now at Manchester Metropolitan University, Manchester, M15 6BH, UK}, 
M.~F.~Cicala$^c$,
T.~Conneely$^d$,  
D.~Cussans$^a$, \\
M.~W.~U.~van~Dijk$^e$,
R.~Forty$^e$,
C.~Frei$^e$, 
E.~P.~M.~Gabriel$^{f,}$\thanks{Now at Nikhef National Institute for Subatomic Physics, Amsterdam, Netherlands}, 
R.~Gao$^g$, 
T.~Gershon$^c$, \\
T.~Gys$^e$, 
T.~Hadavizadeh$^{g,}$\thanks{Now at School of Physics and Astronomy, Monash University, Melbourne, Australia}, 
T.~H.~Hancock$^g$, 
N.~Harnew$^g$, 
T.~Jones$^c$,
M.~Kreps$^c$, \\
J.~Milnes$^d$, 
D.~Piedigrossi$^e$, 
J.~Rademacker$^a$, 
J.~C.~Smallwood$^{g,}$\thanks{Corresponding author's email: jennifer.smallwood@cern.ch}
        }

\usepackage{amsmath}
\usepackage{cite} 
\usepackage{mciteplus}

\begin{document}
\begin{titlepage}
\maketitle
\thispagestyle{empty}

\centering 
$^a$ H.H. Wills Physics Laboratory, University of Bristol, Tyndall Avenue, Bristol BS8 1TL, UK\\
$^b$ University of Bath, Claverton Down, Bath BA2 7AY, UK\\
$^c$ Department of Physics, University of Warwick, Coventry, CV4 7AL, UK\\
$^d$ Photek Ltd., 26 Castleham Road, St Leonards on Sea, East Sussex, TN38 9NS, UK\\
$^e$ CERN,  CH 1211, Geneva 23, Switzerland\\
$^f$ School of Physics and Astronomy, University of Edinburgh, James Clerk Maxwell Building, Edinburgh EH9 3FD, UK\\
$^g$ Denys Wilkinson Laboratory, University of Oxford, Keble Road, Oxford OX1 3RH, UK\\

\begin{abstract}
\noindent
TORCH is a novel time-of-flight detector,  designed to provide charged particle identification of pions, kaons and protons in the momentum range  2--20\gevc over a 9.5\m flight path. 
A detector module,  comprising a 10\mm thick quartz plate,  provides a source of Cherenkov photons which propagate via total internal reflection to one end  of the plate. Here, the photons are focused onto an array of custom-designed  Micro-Channel Plate Photo-Multiplier Tubes (MCP-PMTs) which measure their positions and arrival times. 
The target time resolution per photon is 70\ps which, for 30 detected photons per charged particle, results in a $10-15\ps$ time-of-flight resolution. 
A 1.25\m length TORCH prototype module employing two MCP-PMTs has been developed, and tested at the CERN PS using a charged hadron beam of 8\gevc momentum. The construction of the module, the properties of the MCP-PMTs and the readout electronics are described.
Measurements of the collected photon yields and single-photon time resolutions have been performed as a function of particle entry points on the plate and compared to expectations. 
These studies show that the  performance of the TORCH prototype approaches the design goals for the full-scale detector. 
\end{abstract}

\end{titlepage}

\section{Introduction}
\label{sec:intro}

The TORCH detector (Time Of internally Reflected CHerenkov light) is designed to perform particle identification of low-momentum (2-20\gevc) charged hadrons using time-of-flight. 
The TORCH detector uses prompt Cherenkov photons which are produced when charged particles traverse a 10\mm thick quartz plate,  combining  DIRC-style reconstruction with timing measurements~\cite{Charles:2010at,Brook:2018qdc}. The method  extends the techniques which were pioneered by the BaBar DIRC \cite{Adam_2005} and Belle II iTOP \cite{Abe_2010, Fast_2017} collaborations. Total internal reflection traps a fraction of the photons within the plate, which propagate to focusing optics positioned at the periphery of the plate. Inside the focusing optics, the photons are reflected onto photo-sensitive detectors by a cylindrical mirrored surface. The geometry allows for the Cherenkov angle of emission to  be reconstructed and the time-of-arrival recorded.

The principle of the TORCH time-of-flight (ToF) measurement is detailed in Ref.\,\cite{Brook:2018qdc}.  From  knowledge of the particle entry position in the radiator and the photon hit coordinate,  the  Cherenkov emission angle $\theta_c$ and hence the photon path length in the quartz $L$ can be  calculated by unfolding the multiple internal reflections of the photon. Due to chromatic dispersion in the radiator, photons with different wavelengths propagate at different speeds according to the group refractive index $n_g$ of quartz. The expression $\cos  \theta_c = 1/(\beta n_p)$ relates $\theta_c$ to the phase refractive index $n_p$ which in turn yields  $n_g$ via a known dispersion relation. The time-of-propagation in the quartz and thence the ToF can then be calculated from the inferred group velocity and $L$. The method also requires knowledge of $\beta$, the speed of the charged particle,
which can be calculated from the known particle momentum and measured time  for each particle mass hypothesis in turn. This can be propagated through  subsequent reconstruction stages, allowing the preferred mass  to be selected.

\begin{figure}[ht]
    \centering
    \begin{subfigure}[]{0.49\linewidth}
        \includegraphics[width=\linewidth]{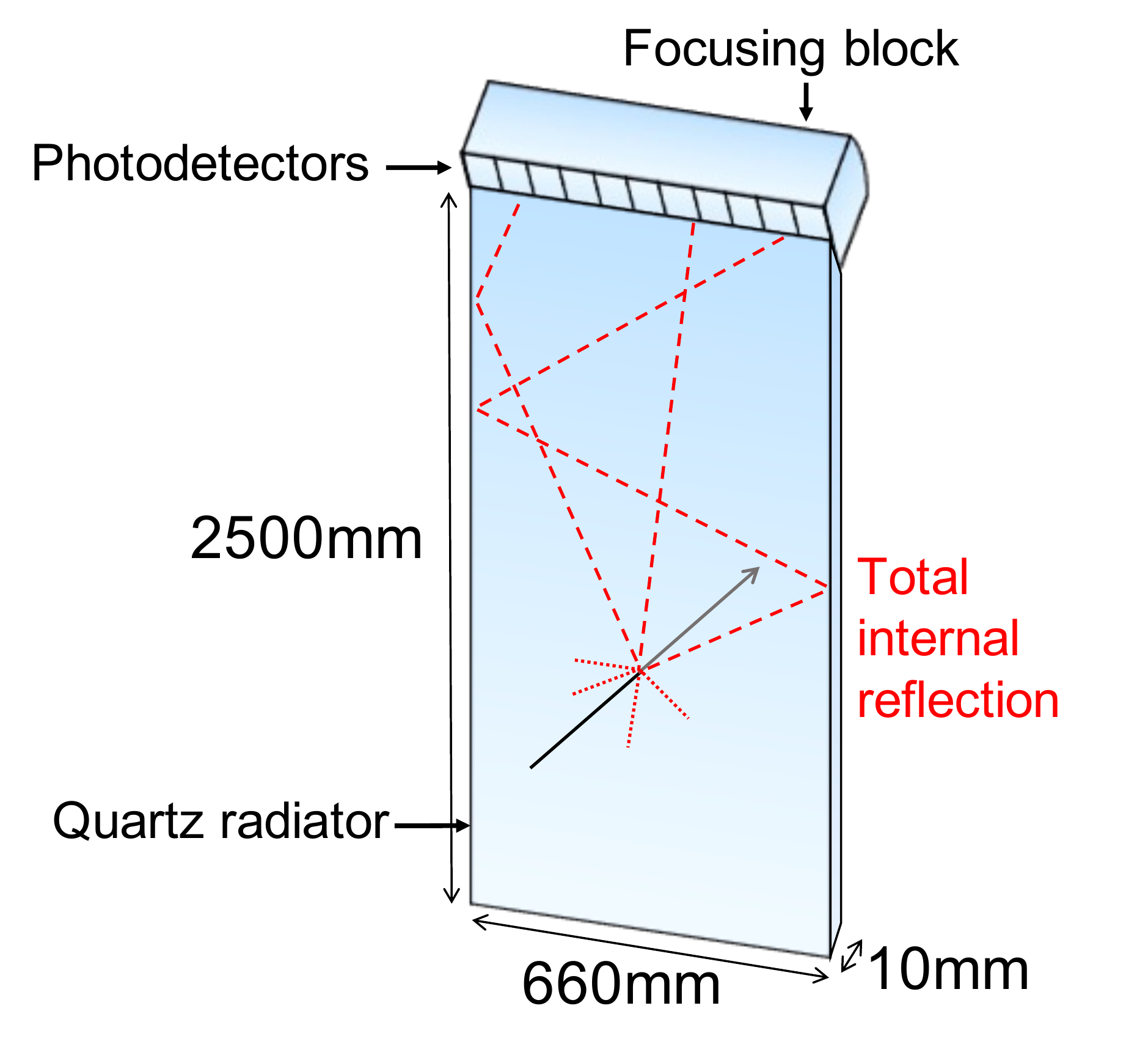} 
        \caption{}
        \label{fig:design_module}
    \end{subfigure}
    \begin{subfigure}[]{0.49\linewidth}
        \includegraphics[width=\linewidth]{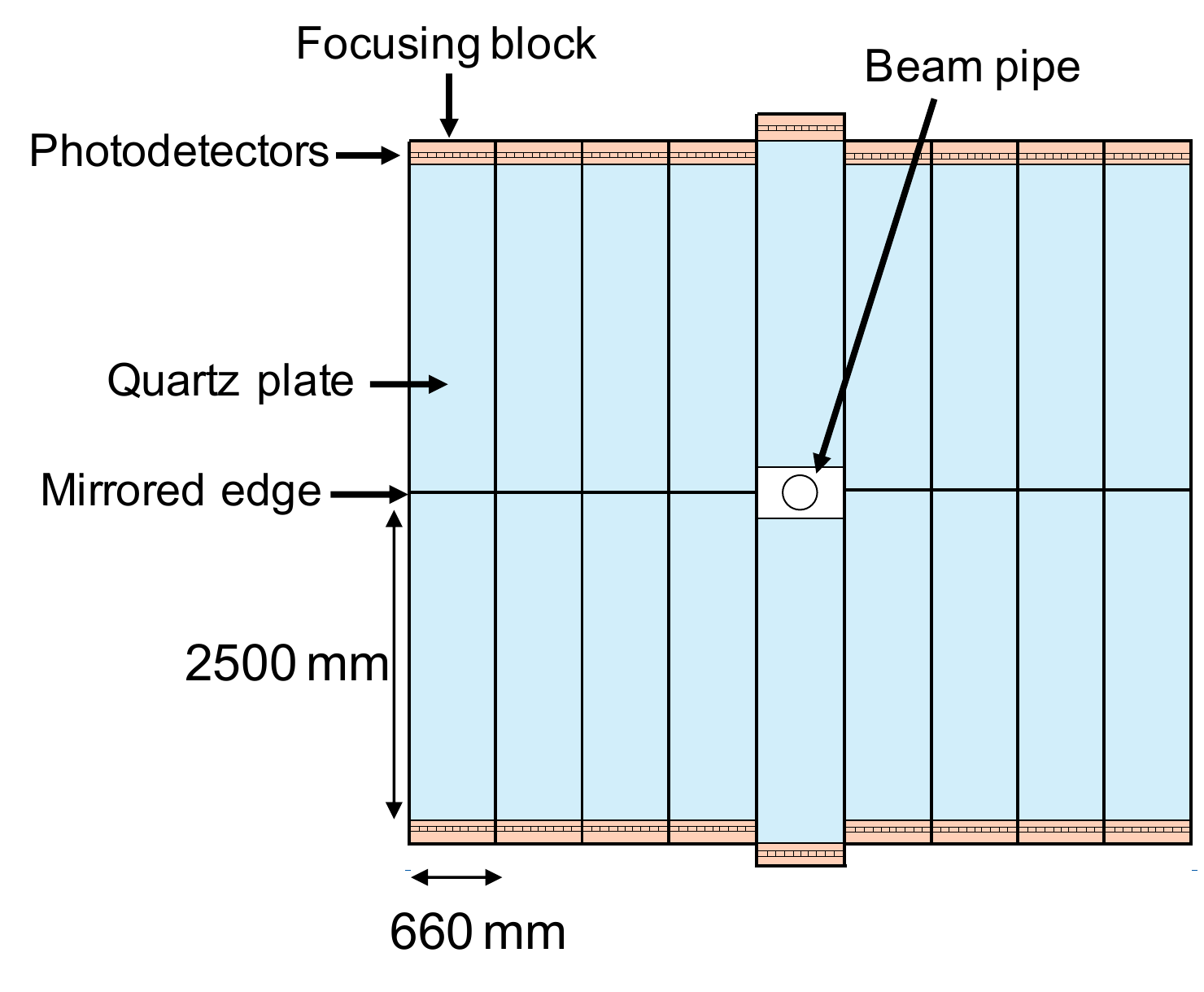}
        \caption{}
        \label{fig:design_detector}
    \end{subfigure}
    \caption{
    TORCH detector design for LHCb Upgrade~II. (a) A single TORCH module with the charged particle path (black line) and the Cherenkov photons (red lines), 
    (b) the full TORCH detector. }
    \label{fig:design_TORCH_module}
\end{figure}

The design of a single full-sized TORCH module is shown in Fig.~\ref{fig:design_module},
with radiator dimensions ${660 \times 2500 \times 10\mmv}$ (width $\times$ length $\times$   thickness). A modular design has been employed to allow coverage across large areas with flexibility  of geometries.
The proposed design for use in the Upgrade~II of the LHCb experiment~\cite{LHCbCollaboration:2776420} is shown in Fig.~\ref{fig:design_detector}. 
Here the TORCH detector is composed of eighteen modules covering an area of $6\m\times5\m$.
Each module will be equipped with eleven Micro-Channel Plate Photomultiplier Tubes (MCP-PMTs). When positioned at a distance of $9.5\m$ from the LHCb interaction point, the ToF difference between $10\gevc$ pions and kaons is $35\ps$, requiring a per-track time resolution of $17.5\ps$ for two standard deviations ($\sigma$) and $11.7\ps$ for $3\sigma$ of separation between mass hypotheses. 
Given $\sim 30$ detected photons per track, the requirement on the single-photon time resolution is chosen to be $\sim 70\ps$~\cite{VanDijk2014}. 
In addition, the photons' time-of-propagation within the quartz increases the time separation between signals from the different species due to the differing Cherenkov angles.

This paper builds upon the work presented in Refs.~\cite{Brook:2018qdc, Bhasin2020}, where a small-scale ($120 \times 350 \times 10 \mm^3$) TORCH demonstrator was tested with both a circular MCP-PMT and a single square MCP-PMT. Both prototype MCP-PMTs were produced by Photek, UK~\cite{Conneely:2015}.  
Here the work concerns the development of a larger, half-length full-width, TORCH module instrumented with two customised Photek MCP-PMTs and readout electronics,   described in Section~\ref{sec:design_prototype}. 
This TORCH demonstrator has been tested in a beam  at the CERN PS.
Section~\ref{sec:simulations} reports the modelling of the demonstrator. The test beam infrastructure is presented in Section~\ref{sec:testbeamsetup}. Section~\ref{sec:dataprocessing} describes the data-driven calibrations made to the data.
In Section~\ref{sec:analysis} 
the photon counting efficiency of the prototype is reported,   determined by comparing yields in data and simulation. 
In addition, the timing resolutions of both single photons and hadron tracks are measured as a function of track position in the radiator, after which a comparison is made to the design goals.
Finally, a summary and the outlook for future work is presented in Section~\ref{sec:conclusions}.

\section{Design of the prototype}
\label{sec:design_prototype}

The tests described in this paper were undertaken with a TORCH prototype module, hereafter referred to as \textit{Proto-TORCH}. The prototype is constructed from a $660 \times 1250 \times 10\mmv$ (width $\times$ length $\times$ thickness) fused silica radiator plate. The plate, which has half the length of a full TORCH module, is glued to a full-sized focusing block with matching width at the top of the plate. 
In addition, specialised mechanics were manufactured  for mounting the MCP-PMTs and the accompanying electronics.
In this section the quartz radiator plate, the focusing block, and the mechanical construction of the support structure holding the optical assembly are reported, together with a description of the MCP-PMTs and readout electronics. 

\begin{figure}[b]
    \centering
    \begin{subfigure}[]{0.45\linewidth}
        \includegraphics[width=\linewidth]{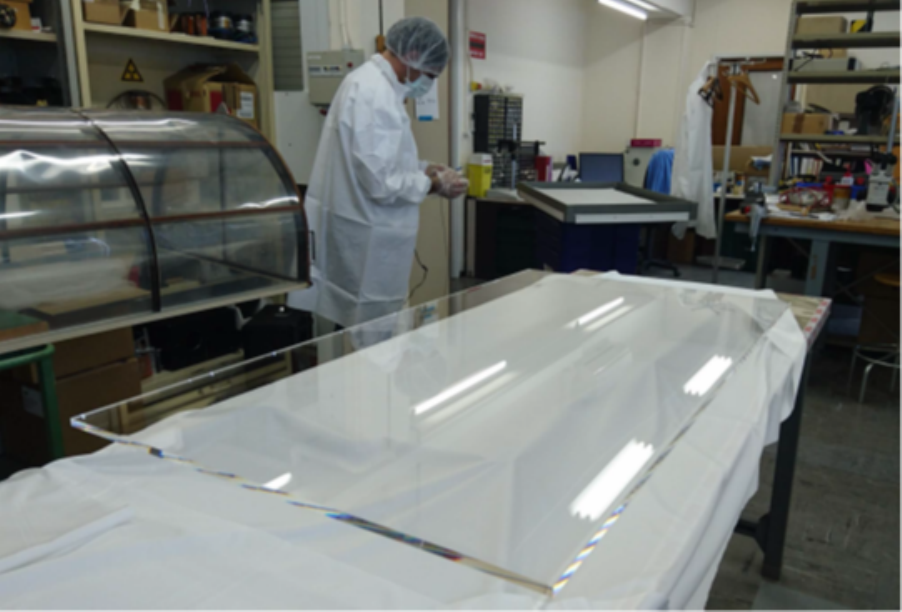} 
        \caption{}
        \label{fig:photo_radiator_plate}
    \end{subfigure}
    \hspace{1.0truecm}
        \begin{subfigure}[]{0.45\linewidth}
        \includegraphics[width=\linewidth]{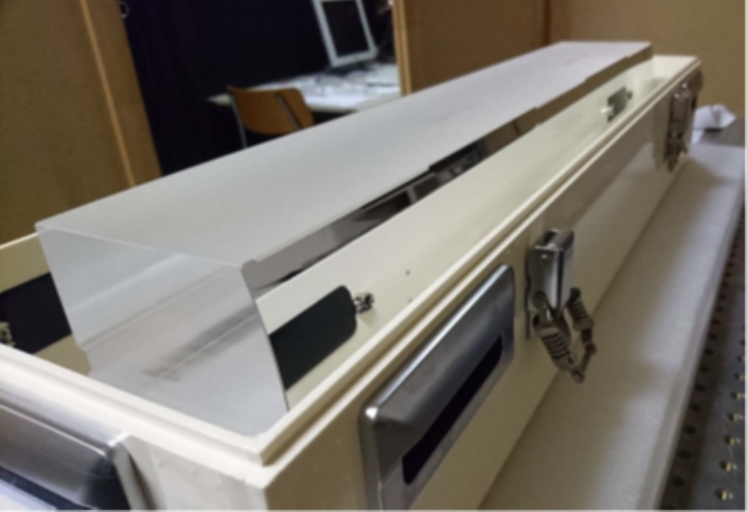} 
        \caption{}
        \label{fig:photo_focusing_block}
    \end{subfigure}
    \caption{
    (a) The manufactured  radiator plate. (b) The focusing block. }
    \label{fig:optics-photos}
\end{figure}

\subsection{Quartz radiator plate and focusing block}
\label{sec:mechanics_quartz}

Photographs of the manufactured radiator plate and focusing block are shown in Fig.~\ref{fig:optics-photos}. All optical components were procured from Nikon Glass\footnote{Nikon Corporation Glass Business Unit, Sagamihara-City, Kanagawa 252-0328 Japan.}. The flat plate surfaces of the radiator plate were polished to a surface roughness of $\sim 0.5\nm$ to minimise surface scattering. With this level of control, simulations show that $90-95\%$ of photons that undergo $\sim 200$ internal reflections are retained\cite{VanDijk:2154410}. 
All side faces of the plate are also polished, including the bottom surface to allow upward reflections of downward travelling photons. 
All edges of the plate are bevelled to 0.1\,mm to avoid sharp edges.

\begin{figure}[t]
    \centering
    \begin{subfigure}[]{0.33\linewidth}
        \includegraphics[width=\linewidth]{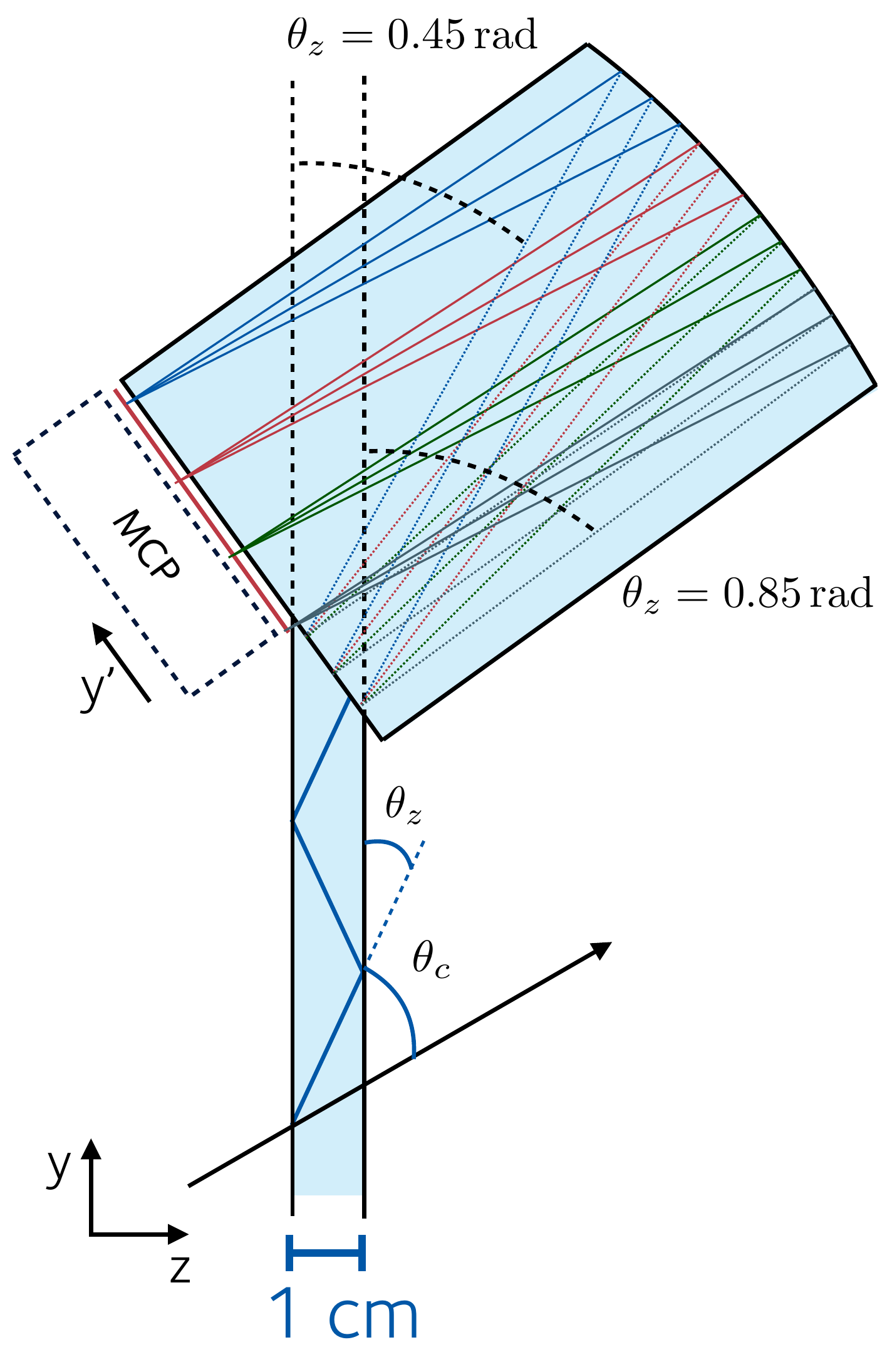}
        \caption{}
        \label{fig:design_focusing_block}
    \end{subfigure}
    \begin{subfigure}[]{0.65\linewidth}
        \includegraphics[width=\linewidth]{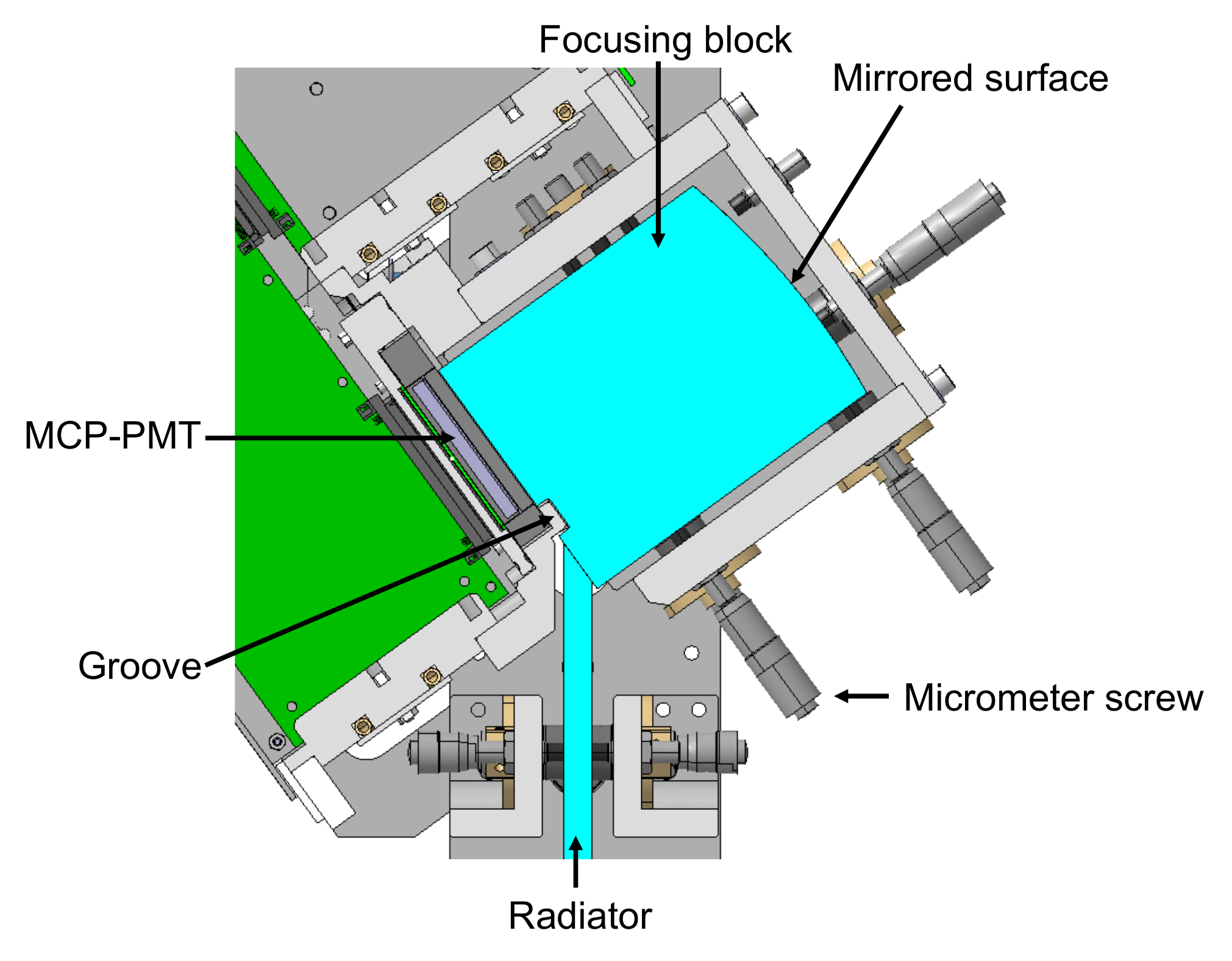} 
        \caption{}
        \label{fig:design_focusing_mount}
    \end{subfigure}
    \caption{
    (a) A schematic representation of the focusing block, not to scale, which shows the translation of the photon's angle of exit from the quartz plate in the vertical plane into the vertical position on the MCP-PMT plane. The charged particle path is shown by the black line with the Cherenkov photons shown by the coloured lines. 
    (b) A technical drawing for the design of the Proto-TORCH mechanics. Detail is given of the focusing-block mounting  showing the groove in the block, the mounting plate, and aligning micrometer screws. }
    \label{fig:design_focusing}
\end{figure}

An illustration of the focusing block is shown in Fig.~\ref{fig:design_focusing_block}. 
The block has a cylindrical mirrored (aluminised) surface with a focal point $5.5\mm$ beyond the exit surface of the block, which is at an angle of $36^\circ$ to the radiator face. 
The MCP-PMTs have 5.0mm-thick input windows which are spaced 0.5mm from the focusing block with an air gap in between (to avoid gluing), such that the focal point coincides with the photo-cathode plane.
Photons propagate at an angle $\theta_z$ relative to the radiator face and upon entering the block are reflected before reaching the MCP-PMT plane. 
The curvature of the mirror is such that photons emitted with the same angle $\theta_z$ from any point of the radiator are reflected onto the same vertical ($y^\prime$) coordinate on the MCP-PMT plane. 
The acceptance of photons on the detector plane is limited to photons that enter the focusing block at angles in the range $0.45 < \theta_z < 0.85$\,rad.
The reflectivity of the mirrored surface  has been measured to be above 85\% 
over the photon energy range of interest (and is further discussed in Sec.~\ref{sec:simulations}).
The side faces are also critical reflective surfaces and as such are polished.
The top and bottom faces are kept rough   so light might be diffused    outside the block.  Whilst this prevents correlated reflections, associated scattering does result in some out-of-time background.  

The main structural support for  the focusing block  is via a mounting plate which is located inside a groove in the block, and this holds the block onto an outer frame, shown in Fig.~\ref{fig:design_focusing_mount}. The groove can contribute towards multiple reflections within the focusing block and, to mitigate this, the groove surface is is kept rough and painted black.
At each end, there is an adjustable micrometric screw to hold the block laterally.

The focusing optics and radiator plate were produced separately, and were bonded together at CERN with 
Pactan 8030 silicone-based adhesive  
which has good transmission characteristics up to 6\,eV in photon energy\cite{CastilloGarcia:2202368} (further discussed in Sec.~\ref{sec:simulations}). 
The advantage of this adhesive is that it does not set rigidly and is suited to  disassembly if the need arises.

\begin{figure}[!htb]
    \centering
    \begin{subfigure}[m]{0.69\linewidth}
        \includegraphics[width=\linewidth]{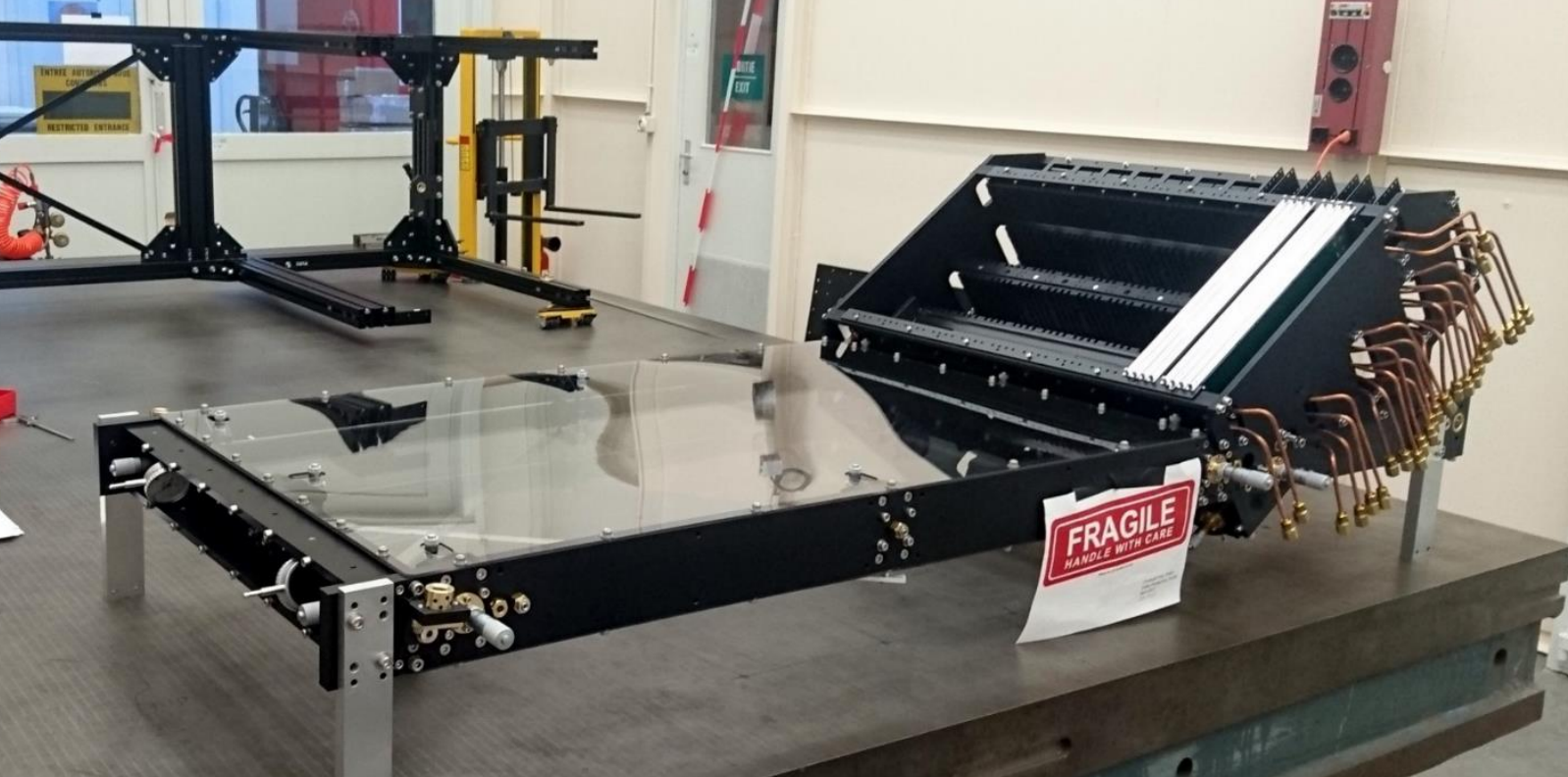}
        \caption{}
        \label{fig:mechanics_support_frame}
    \end{subfigure}
    \begin{subfigure}[m]{0.59\linewidth}
        \includegraphics[width=\linewidth]{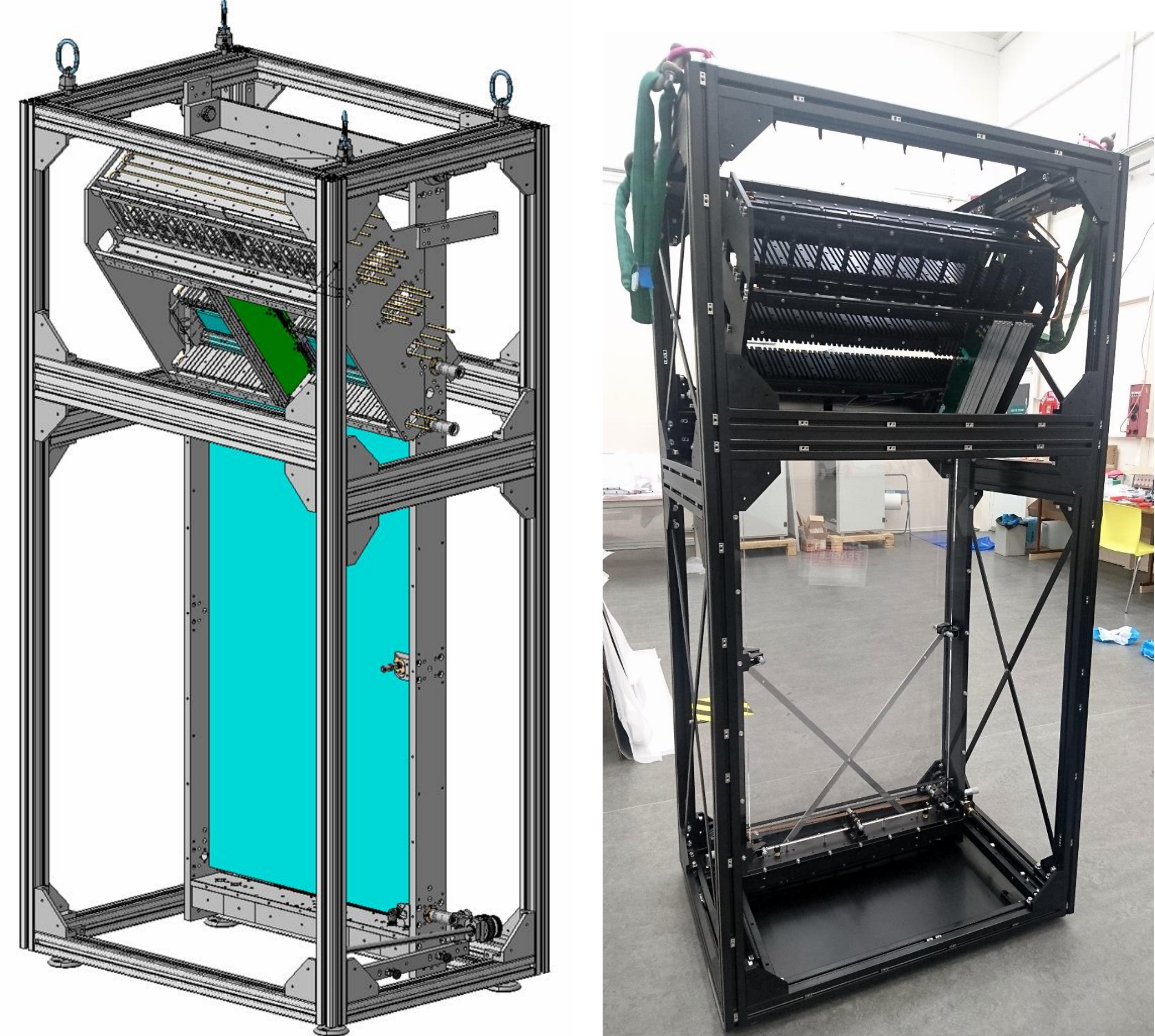}
        \caption{}
        \label{fig:mechanics_full_assembly}
    \end{subfigure}
    \caption{
    (a) The optical support frame for the quartz radiator, lying horizontally; the micrometer adjustments are visible at the bottom left, and the  support structure  for the MCP-PMTs and electronics, on the right. 
    (b) The fully assembled prototype in its casing  (left) the CAD model, and (right) after construction.}
    \label{fig:mechanics_prototype}
\end{figure}

\subsection{Mechanical frame} 
\label{sec:mechanics_case}

The optical components are mounted in a rigid anodised aluminium mechanical frame, as shown in Fig.~\ref{fig:mechanics_support_frame}. 
The focusing block and the quartz plate assembly is finely located via micro-metric screws providing (almost) point contact to the side surfaces of the plate. As these surfaces are reflective-critical, the contact  area is minimised by design. The radiator plate is supported from the bottom by two micro-metric screws and two additional spring loaded stoppers. These latter elements   compensate for thermal expansion of the aluminium frame. All point contacts are terminated with black bumpers of dimension $\sim$1.7\,cm$^2$, and made of polyoxymethylene (POM). 

To bond the optical components,  the frame acts as a jig for applying the optical adhesive between  the quartz plate and the focusing block. With the implementation of micro-metric screws, the radiator plate is first translated near to the focusing block. Then, by suitable fine-adjustment of the screws, this allows  a thin and carefully-controlled coating of glue of thickness 55\,$\mu$m, which minimises transmission light-loss.

A mechanical structure that houses the MCP-PMTs and all the readout electronics is attached to the optical assembly described above. This structure implements so-called ``comb plates'', which allow the accurate location of the electronics boards to directly couple to the photo-detectors. 
The readout boards are fitted with metallic plates, coupled with thermal paste, to adequately transfer heat to the comb plates. Copper pipes (visible in Fig.~\ref{fig:mechanics_support_frame})  are embedded in these combs, to provide liquid cooling and to evacuate the heat. Note that the liquid cooling was not used during this test beam campaign as only two MCP-PMTs were used, but the mechanics are designed to allow the system to be extended to the full complement of eleven MCP-PMTs used in a full-scale module. 

The complete assembly: optical components, aluminium frame, photodetectors and electronics, shown in Fig.~\ref{fig:mechanics_full_assembly}, is housed finally within a large and rigid light-tight crate. The assembly is mounted at a rotating point which is  located at the top. This allows a tilt  around the horizontal $x$-axis (perpendicular to the beam direction) and   hence a variation of the incident particle angle through the radiator. A mechanism  at the base ensures a smooth and precise rotation of the assembly up to $10^\circ$.

\subsection{Micro-Channel Plate PMTs}
\label{sec:MCPs}

During the beam tests, the Proto-TORCH module was instrumented with a pair of Photek-developed MCP-PMTs, the final devices of a 3-stage  R\&D plan~\cite{Conneely:2015}. The two MCP-PMTs were located adjacent to the upper corner of the radiator plate; 
their associated electronics readout boards are visible in Fig.~\ref{fig:mechanics_prototype}. 

\begin{figure}[t]
    \centering
    \begin{subfigure}[m]{0.40\linewidth}
        \includegraphics[width=\linewidth]{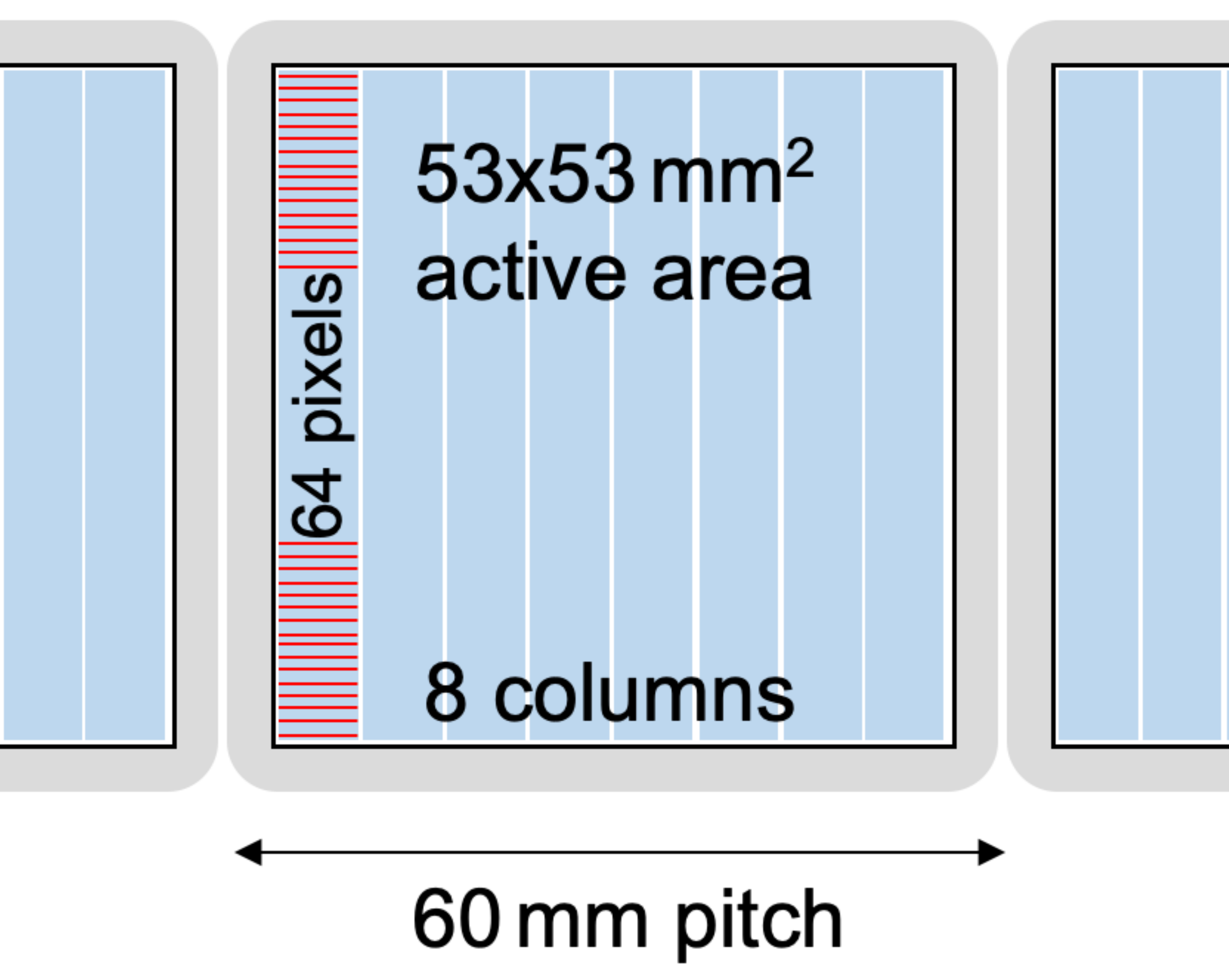}
        \caption{}
        \label{fig:MCPs_pixel}
    \end{subfigure}
    \begin{subfigure}[m]{0.58\linewidth}
        \includegraphics[width=0.52\linewidth]{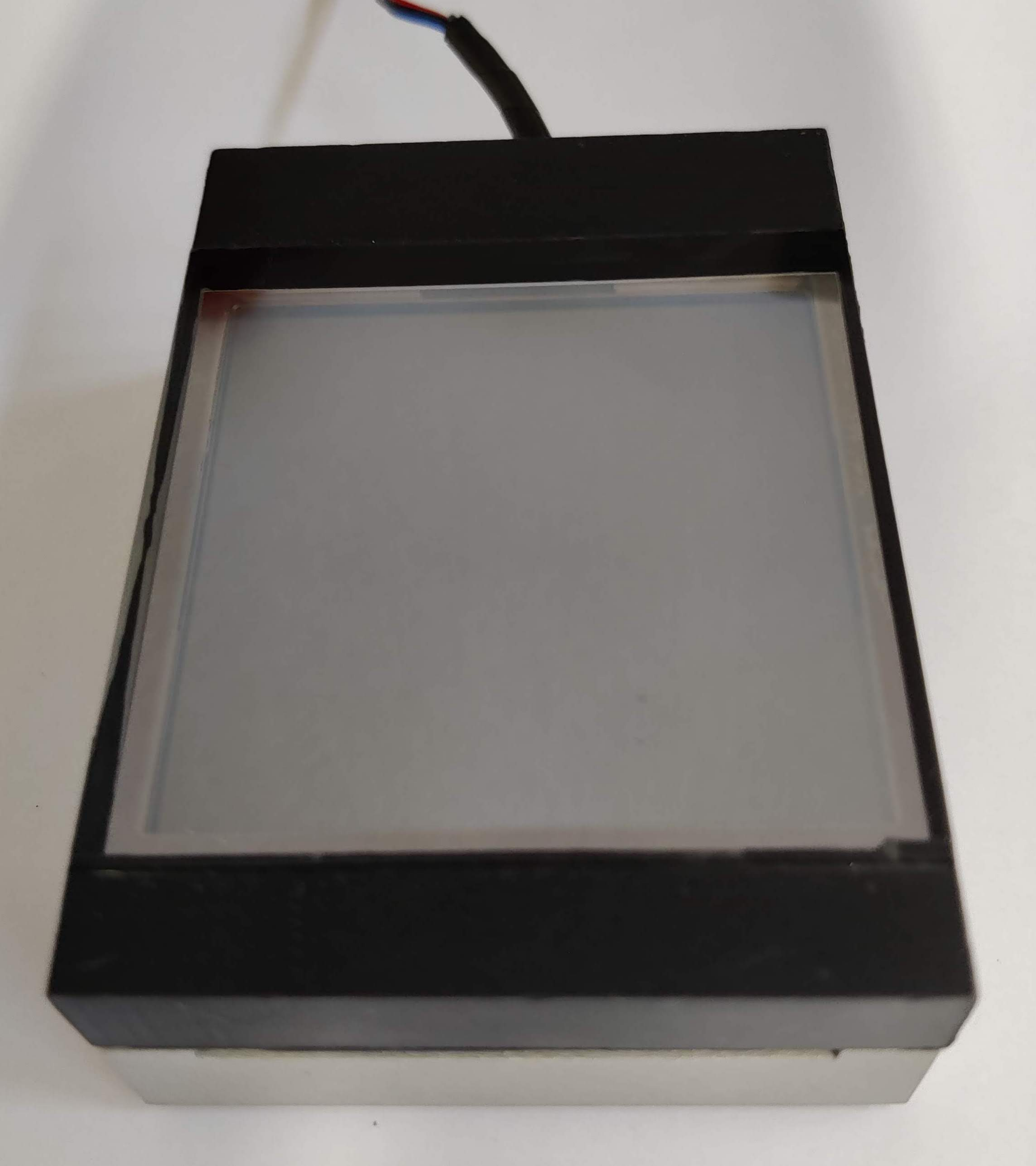}
        \includegraphics[width=0.42\linewidth]{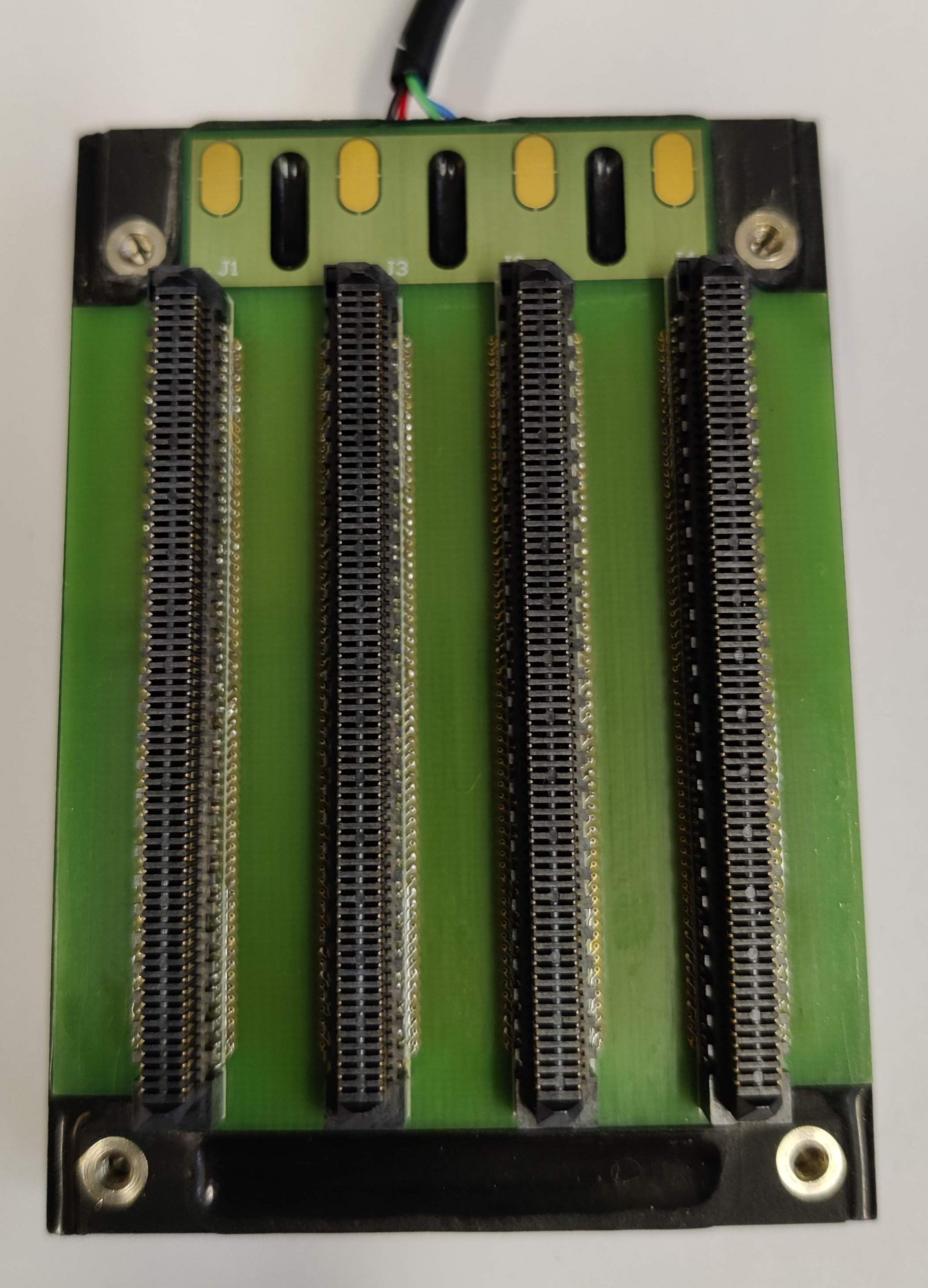}
        \caption{}
        \label{fig:MCPs_connectors}
    \end{subfigure}
    \caption{
    (a) The MCP-PMT pixel layout where the $64\times64$ pixels are grouped to form 8 columns and charge sharing is used to increase the effective granularity to 128 rows. 
    (b) The MCP-PMT front surface and the connectors on the rear side.} 
    \label{fig:MCPs}
\end{figure}

Both MCP-PMTs have an intrinsic pixelisation of $64\times 64$ pads over a $53\times53$\,mm$^{2}$ active area, within a $60\times60$\,mm$^2$ physical footprint. 
The MCP-PMT anode pads are connected directly to an external printed circuit board (PCB)  using anisotropic conductive film. The MCP-PMTs are read out with a granularity of $8\times64$ pixels, achieved by electronically grouping channels on the PCB  in the horizontal direction,  and seen schematically in Fig.~\ref{fig:MCPs_pixel}. The front surface and back-plane connectors of the MCP-PMT  are shown in Fig.~\ref{fig:MCPs_connectors}.

The MCP-PMTs are designed to withstand an integrated anode charge of $5\,{\rm C}/{\rm cm}^{2}$ \cite{GYS2017156} by utilising an atomic layer deposition (ALD) coating on the micro-channel plates. 
The electron avalanche within the MCP-PMT deposits charge on a resistive layer inside the PMT, which is capacitively coupled to the anode pads via an insulating layer. 
Since the avalanche of charge produced by a single photon in the MCP typically causes hits on multiple channels, the centroid of these hits allows the spatial and timing precision of the photon arrival  to be improved using a charge-weighting technique in the vertical direction. This achieves an effective $8\times128$ granularity, 
which provides a sufficient angular precision on the reconstructed Cherenkov angle to meet the TORCH single-photon timing requirements.

The single-photon timing resolution requirement can be realistically broken down into 50\,ps from the MCP-PMT and readout, and 50\,ps from the time-of-propagation resulting from the finite pixel size. The single-photon time resolution of a 64 $\times$ 64 pixel Photek MCP-PMT has been measured to be $(47.5 \pm 0.7)\ps$ in the laboratory\,\cite{CICALA2022166950}, including the contribution from the customised electronics-readout system, described in the section below.

\begin{figure}[t]
    \centering
    \begin{subfigure}[m]{0.49\linewidth} 
        \includegraphics[width=\linewidth]{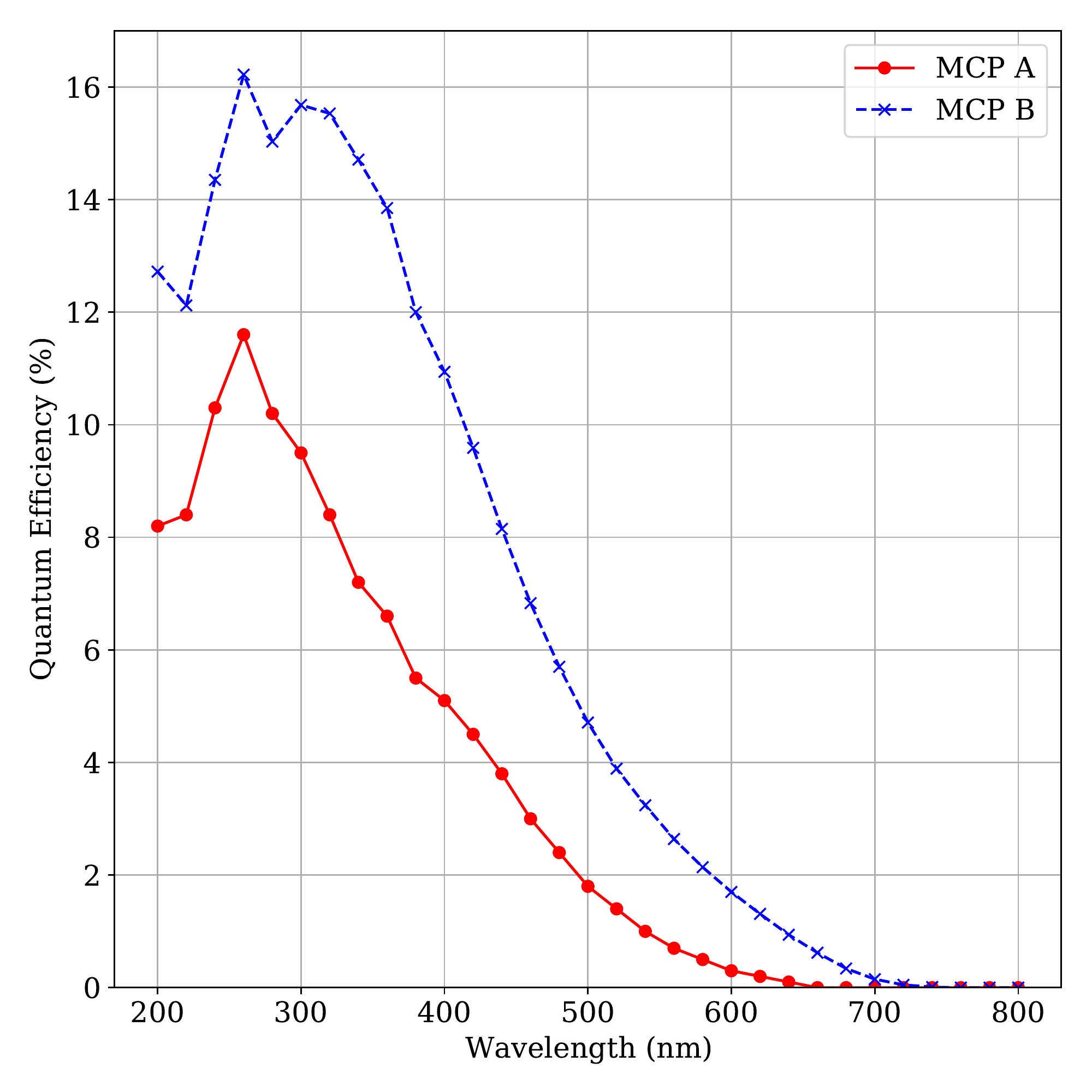} 
        \caption{}
        \label{fig:MCP_QE}
    \end{subfigure}
    \begin{subfigure}[m]{0.49\linewidth} 
        \includegraphics[width=\linewidth]{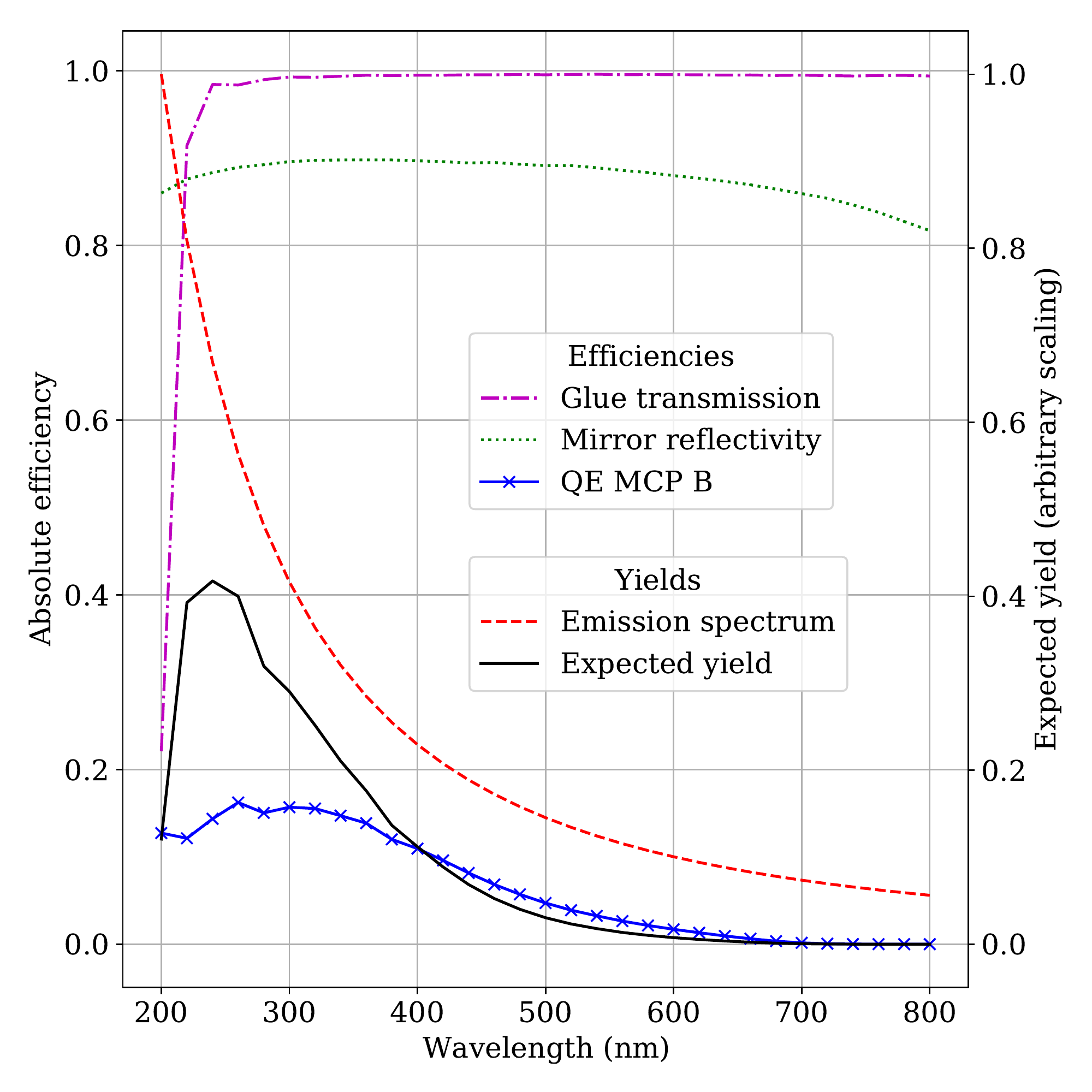} 
        \caption{}
        \label{fig:eff_wavelength}
    \end{subfigure}
    \caption{(a) The measured quantum efficiencies as a function of wavelength of the MCP-PMTs instrumented on Proto-TORCH, measured at CERN.
    (b) The absolute efficiencies of glue transmission, mirror reflectivity and photo-electron conversion (for MCP\,B) as a function of wavelength. The photon emission spectrum and expected photon yield due to the efficiency sources are also shown, both with arbitrary scaling.
    }
    \label{fig:alleff}
\end{figure}

The two MCP-PMTs under test (hereafter labelled MCPs `A' and `B') are  characterised by their quantum efficiency (QE), multiplication factor and charge avalanche point-spread function (PSF). 
The QEs of both MCP-PMTs were measured in the laboratory prior to the beam tests in October 2018, and the performance differs for MCPs A and B, as shown in Fig.~\ref{fig:MCP_QE}. 
The higher QE of MCP\,B resulted in a greater photon yield for that tube.
The collection efficiency of the MCP-PMTs, defined as the probability for a converted photoelectron to be multiplied through the MCP chain, is quoted by Photek to be around $65\%$. 
The gain in each MCP-PMT is estimated to be around $1.25 \times 10^6$ electrons for MCP\,A and $1.35 \times 10^6$ electrons for MCP\,B, based on the calibrated bleeder-chain HV settings. 
The gain was measured in the laboratory and was found to vary over the MCP-PMT plane: up to 17\% across MCP B, but as high as 50\% across MCP A. 
The insulating layer has a thickness of $0.3\mm$ on MCP\,A which results in a PSF FWHM at the anode pads of $1.30 \pm 0.13\mm$~\cite{Bhasin2020}. 
MCP\,B has a $0.5 \mm$ thick insulating layer, with a PSF FWHM of $1.90 \pm 0.06 \mm$. 

\subsection{Electronics and trigger}
\label{sec:electronics}

The signals from the MCP-PMTs are read out using customised electronics boards
\cite{Gao:2021abc} that use chip-sets developed for the ALICE TOF detector~\cite{Akindinov:2004gf}. A modular electronics system for a small-scale prototype providing up to 256 channels per system in a 4-board $\times$ 64-channel arrangement has been reported previously\,\cite{Gao:2016brz} and the Proto-TORCH electronics design evolves  from this. 

A set of four NINO ASIC chips~\cite{Anghinolfi:2004gg} are located on the so-called {\it NINO board}.
Four  NINO boards are connected to a single MCP-PMT, each through one of the four connectors mounted on the MCP-PMT rear surface. 
Each NINO channel amplifies the input signal and outputs a square pulse, for which the leading edge corresponds to the time that the leading edge of the charge deposit crosses the threshold; the length indicates the Time-Over-Threshold (TOT). 
The NINO charge thresholds are estimated to be 20-30 fC over all channels. This is based on a measurement of four channels from two NINO boards at a threshold setting of 100\,mV (1.65V/1.55V) made previously in Ref.~\cite{CastilloGarcia:2202368}. 
A single NINO chip comprises 32-channels, giving a total of 128 channels per NINO board.

The two edges of the NINO output pulse are then time-stamped by a High Performance Time to Digital Convertor (HPTDC) ASIC~\cite{Mota:2000abc}, located on the {\it HPTDC board}. A single HPTDC board comprises two 32-channel HPTDCs.
The HPTDC time digitisation was set throughout the beam tests close to 100\,ps binning (256 bins in 25\,ns), which provides a small contribution to the final measured Proto-TORCH timing resolution of 100/$\sqrt{12}$\,ps.

A {\it Readout board} collects digitised data from four HPTDC boards and formats them into Ethernet packets to be sent to a DAQ PC.  
The data include the leading edge time of the signal and the TOT. The TOT is correlated to the charge deposited in the pixel and is used to correct for time-walk of the signal.
The non-linearities in the HPTDC binning are also corrected offline, with details given in Sec.~\ref{sec:dataprocessing}.

The footprint of the new printed circuit boards has conformed to IEC standards, namely a 3U height of 100\,mm with a total stack height of   front and rear boards to fit within the dimensions of a standard  9U crate.
This ensures no interference of the electronics with the siting of the radiator plate.

\begin{figure}[t]
    \centering
    \begin{subfigure}[m]{\linewidth}
        \includegraphics[width=\linewidth]{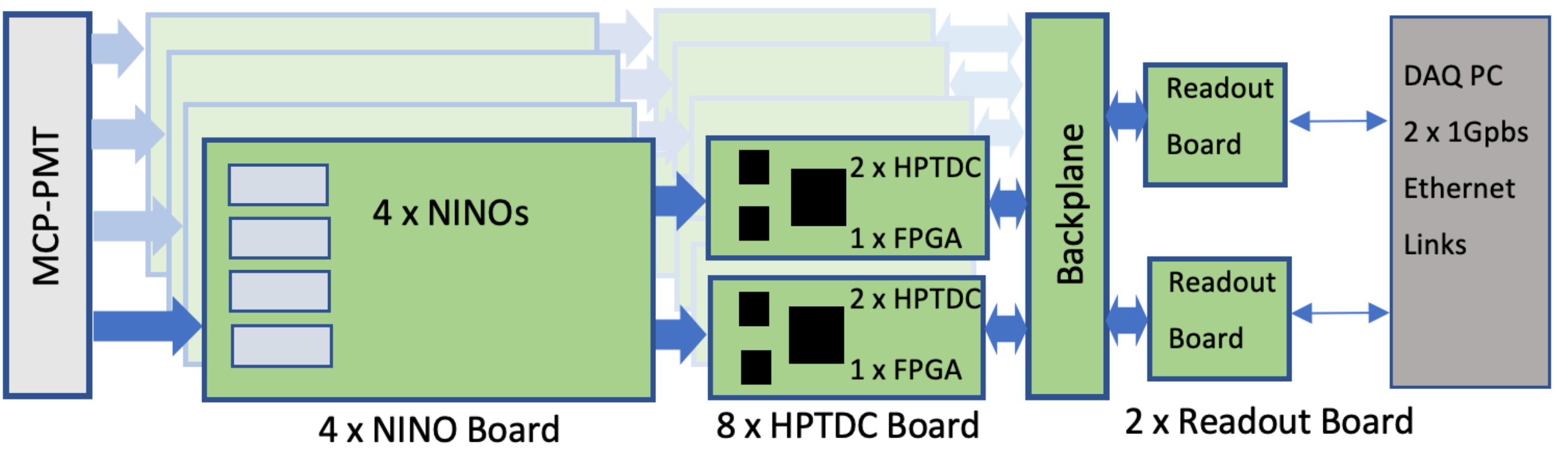}
        \caption{}
        \label{fig:TORCH-electronics-schematics}
    \end{subfigure} \\
    \begin{subfigure}[m]{\linewidth}
    \centering
        \includegraphics[width=0.7\linewidth]{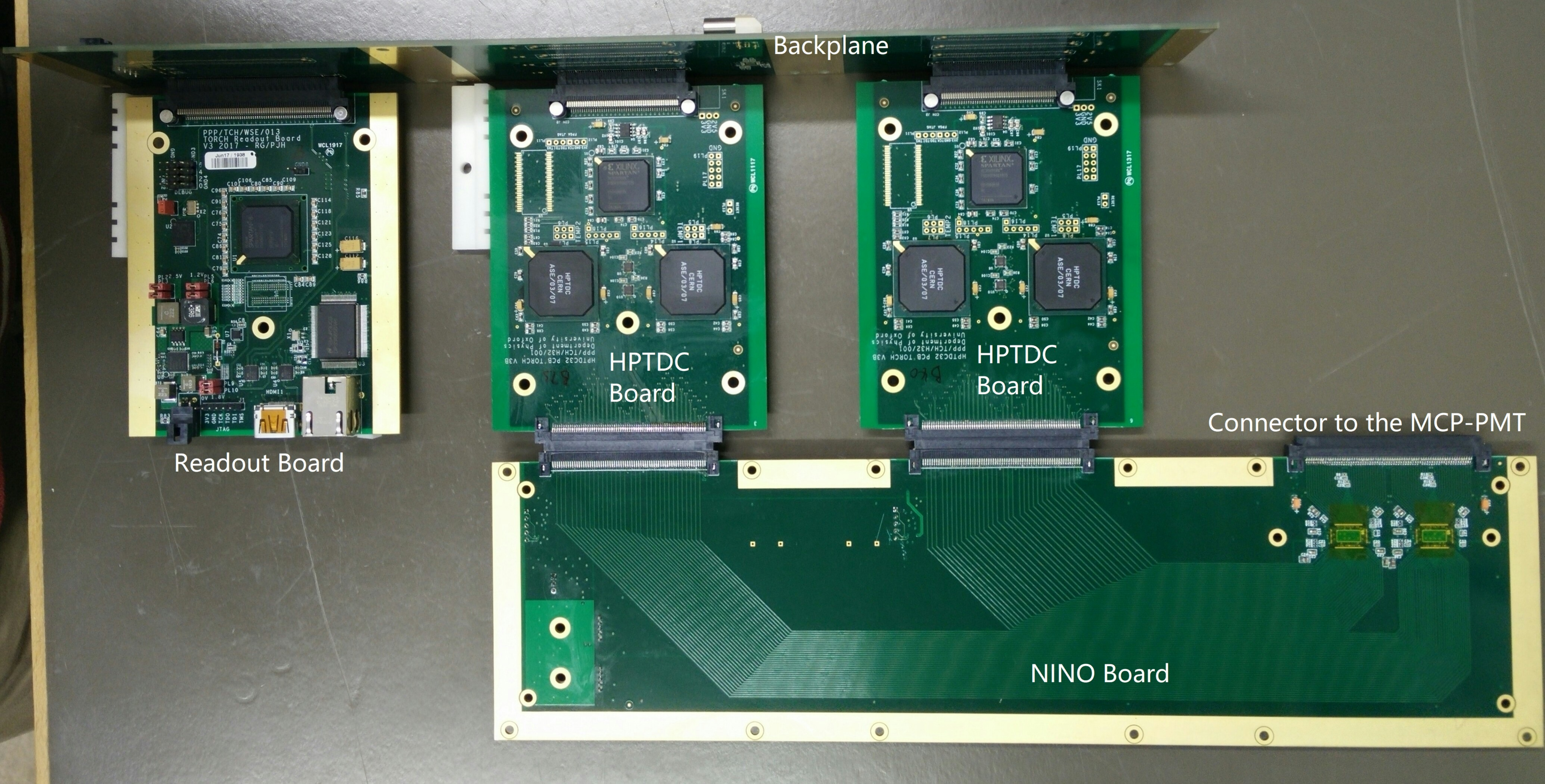}
    \caption{}
    \label{fig:TORCH-electronics}
        \end{subfigure}
            \caption{(a) A schematic and (b) a photograph of the TORCH electronics readout boards. }
    \label{fig:electronics}
\end{figure}

Figure~\ref{fig:electronics}  shows the data-flow of the electronics system and a photograph of the system for a single NINO board, showing also the connector to an MCP-PMT. The Proto-TORCH modularity has each 8 $\times$ 64-channel MCP-PMT served by four NINO boards, eight HPTDC boards, and two Readout boards joined together by a backplane, with the system read out by two 1\,Gbps Ethernet links. 
The Ethernet links are connected to a commercial switch with 10\,Gbps uplink ports, where the DAQ PC is connected. The switch controls the routing, traffic control and buffering, which allows simultaneous readout of all readout boards within the system. 

The readout board communicates with an AIDA Trigger Logic Unit (TLU)\cite{Baesso_2019} using the EUDAQ2 framework \cite{Liu_2019}, which provides a trigger and synchronisation with additional beamline counters. The TLU distributes clock,  trigger and enable signals to the readout boards in the system. Once data-taking starts, the HPTDC measures the photon arrival time and TOT when a trigger signal is received. The readout boards  raise a busy signal to the TLU if a maximum trigger rate is reached. 

\section{Efficiencies and simulation}
\label{sec:simulations}

Monte Carlo simulation is used to provide expectations of the single-photon arrival-time distributions and of the photon yields. In addition, the external contribution towards the time resolution due to the beam profile used in the test beam (described in Sec.~\ref{sec:testbeamsetup}) is estimated using simulated events, 
where the unknown event-by-event beam-particle entry points result in small deviations in the path length of Cherenkov photons. 

The simulation is generated using Geant4~\cite{Agostinelli2003, Allison2006} and takes as input a charged particle of known species, momentum, and entry position on the radiator. 
The profile of the beam entry position is modelled as a 2D Gaussian with widths $2.73\mm$ in $x$ and $2.01\mm$ in $y$, based on measurement in Ref.~\cite{Bhasin2020}. 
The production of Cherenkov radiation is modelled according to the Frank-Tamm relation~\cite{FrankTamm}, with the emission spectrum inversely proportional to the wavelength squared. 
The photons are propagated through a modelled radiator and focusing block, with refractive indices implemented from the manufacturer's specifications. 
The surfaces of the radiator and focusing block are modelled as dielectric-dielectric boundaries, with properties to match the surface types as described in Sec.~\ref{sec:mechanics_quartz}. 
The transmission of photons through the glue that is used to optically couple the radiator and focusing block is accounted for, which imposes a sharp cut-off at low wavelengths around 220\nm. 
The mirrored surface of the focusing block has a wavelength-dependent reflectivity which plateaus in the middle of the range.
The air-gap between the focusing block exit surface and the MCP-PMT window is also included. 
The photon emission spectrum, efficiencies from different sources, and the resulting expected photon yields are shown as a function of photon wavelength in Fig.~\ref{fig:eff_wavelength}.  

Further, the response of the MCP-PMT and NINO chip is modelled. 
The probability of photon conversion in the photocathode is modelled according to the measured QE, which peaks in the wavelength range $250-300\nm$.
The behaviour of the photo-electron through the Micro-Channel Plates is determined by the collection efficiency and the probability of back-scattering, which can lead to spatial displacement and time delay of the charge avalanche. A back-scattering model has been determined based on data taken during laboratory tests of the MCP-PMTs.  
The gain amplification of an individual  photo-electron is randomly sampled from a model developed from measurements of the gain made in laboratory tests. 
The charge avalanche is spread according to the PSF of each MCP-PMT, and a Gaussian integral of the charge-spread is performed to determine which pixels receive sufficient depositions to be above the NINO threshold, and thus have a hit recorded. 
The same clustering algorithm  used in the data analysis (described in  Sec.~\ref{sec:dataprocessing}) is used to cluster the hits into reconstructed photons.

\section{Test beam setup}
\label{sec:testbeamsetup}

\begin{figure}[t]
    \centering
    \begin{subfigure}[m]{\linewidth}
        \includegraphics[width=\linewidth]{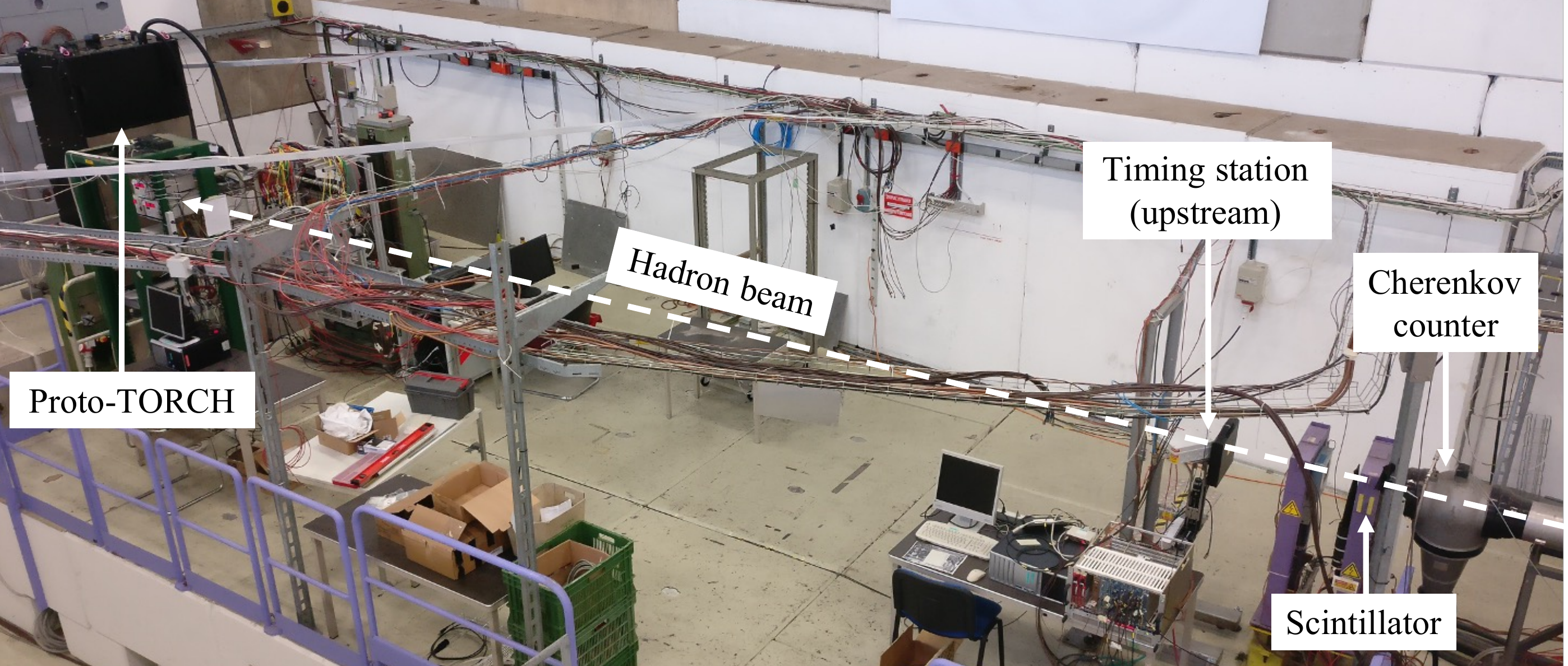} 
        \subcaption{}
        \label{fig:testbeam_area}
    \end{subfigure}
    \begin{subfigure}[m]{\linewidth}
        \includegraphics[width=\linewidth]{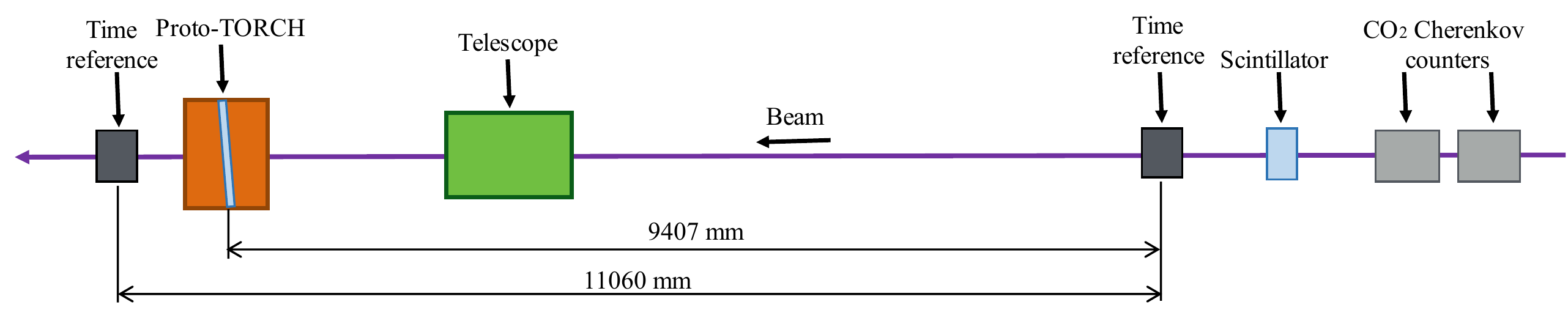}
        \caption{}
        \label{fig:testbeam_schem}
    \end{subfigure}
    \caption{(a) Views of the T9 test beam area along the beam direction. 
    The beam trajectory has been overlaid and travels from right to left. One of the Cherenkov counters, the telescope and the downstream timing station are not visible in the photograph. 
    (b) The layout of the components of the test beam, with the beam travelling from right to left. 
    }
    \label{fig:testbeam}
\end{figure}

The beam test was performed at the T9  area of the CERN PS using a nominal 8\gevc momentum beam composed of $(47.3\pm1.5)\%$ pions and $(52.7\pm1.5)\%$ protons~\cite{Hancock:thesis}. 
The beam-line infrastructure was similar to that of previous test beam campaigns with the smaller module\cite{Bhasin2020}.
The beam-line components are illustrated and are visible in the photograph of the beam test area in Fig.~\ref{fig:testbeam}. 

Two \textit{timing stations} provided  a time reference and were positioned 11\m apart, with Proto-TORCH positioned 1.65\m upstream of the downstream station. Each timing station comprises a $8\times8\times100$\mmv borosilicate bar, coupled to a single-channel MCP-PMT to provide accurate timing, and a pair of crossed scintillators. 
The  borosilicate bar was located relative to the beam such that Cherenkov light produced by traversing particles was oriented directly  towards the MCP-PMT. 
The crossed scintillators, each connected to a PMT, had an overlap of $8\times8$\mma perpendicular to the beam. 
The trigger required a signal in both PMTs of the downstream timing station, which served to narrow the beam spatial definition and ensured the triggered particles passed  through the borosilicate finger.  
The timing signals from both timing stations were injected into the Proto-TORCH readout electronics, eight channels per MCP-PMT, to allow straightforward synchronisation of the data,  which also meant that the timing signals were digitised using the 100\,ps binning of the HPTDC chips.  
This scheme had the slight disadvantage that several pixels of the  MCP-PMT were not read out, to be replaced by data from the time-reference stations.  

A pair of CO$_2$-filled threshold Cherenkov counters were located upstream of the first timing station to discriminate pions from protons in the beam.  These enabled a method of particle identification independent of the ToF provided by the timing stations.
In addition, a large scintillator coupled to a single-channel PMT was used in conjunction with signals from both Cherenkov counters to reduce random noise. 
An EUDET/AIDA pixel beam telescope~\cite{Rubinskiy2014} was used to measure the beam profile~\cite{Bhasin2020}, and some of its scintillating layers were used in the trigger definition.

In the test beam, the TORCH light-tight crate was mounted on a translation table, allowing free movement in both directions perpendicular to the beam direction. The radiator was tilted to $5^{\circ}$ with respect to the vertical in order to improve the light collection efficiency from incident charged particles. 

\begin{figure}
\CenterFloatBoxes
\begin{floatrow}
\ffigbox
  {
  \includegraphics[width=0.6\linewidth]{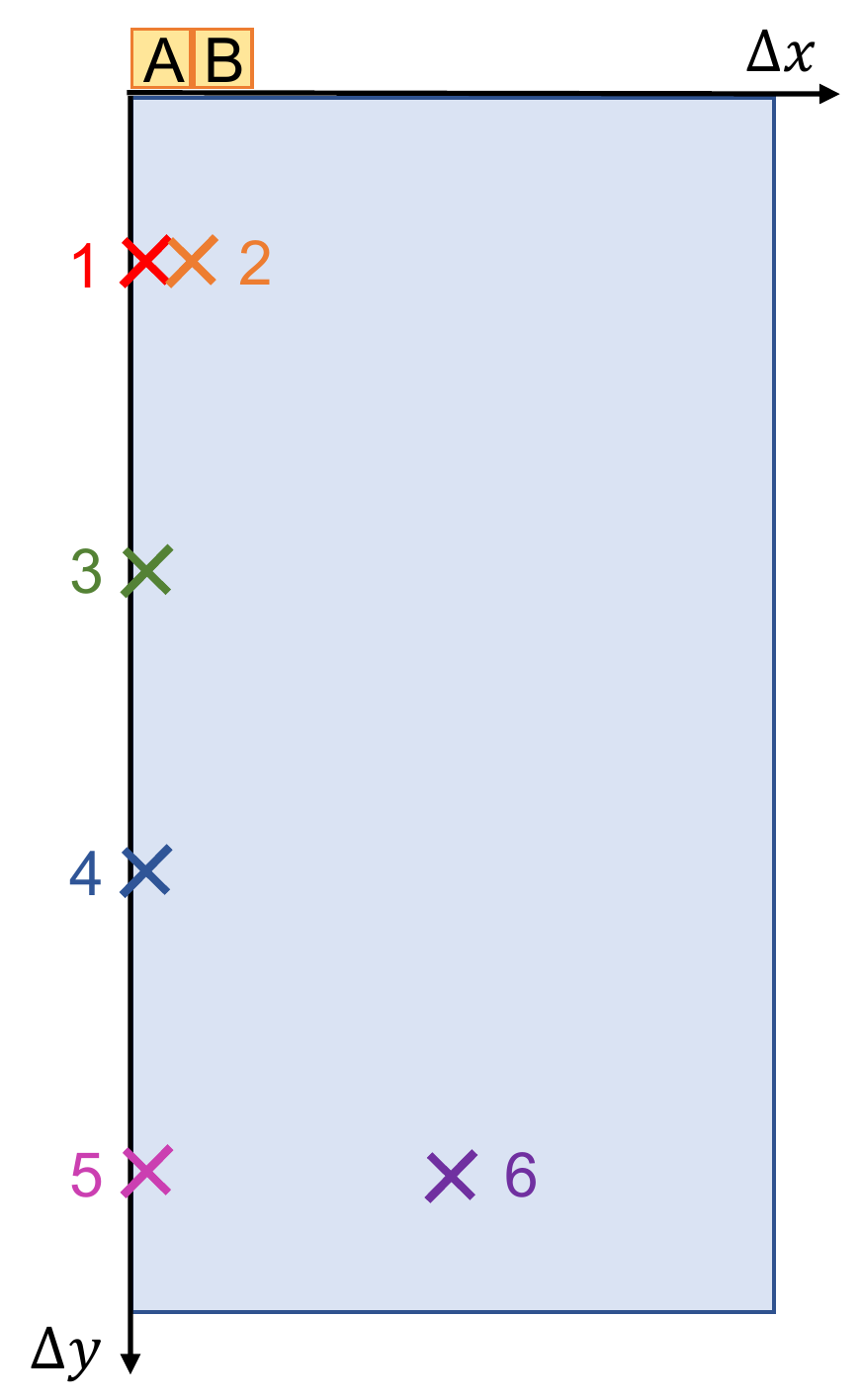}
  }
  {
  \caption[Schematic of Proto-TORCH track positions]{\label{fig:torchpos} Proto-TORCH track positions, with coordinates given in Table~\ref{tab:positions}. }
  }
\killfloatstyle
\ttabbox
  {
  \begin{tabular}{ |c|c|c| } 
\hline
Position & $\Delta x$ (mm) & $\Delta y$ (mm) \\ \hline 
1 & 5 & 175 \\
2 & 60 & 175 \\
3 & 5 & 488.6 \\
4 & 5 & 801.9 \\
5 & 5 & 1115.4 \\
6 & 330 & 1115.4 \\
\hline
\end{tabular}
  }
  {
  \caption[Positions of tracks in Proto-TORCH]{\label{tab:positions} Positions of tracks in Proto-TORCH with the origin taken at the upper corner of the radiator closest to the MCP-PMTs.}
  }
\end{floatrow}
\end{figure}

The performance of Proto-TORCH was studied at a total of six beam-entry positions,  as indicated in Fig.~\ref{fig:torchpos}. The positions relative to the upper left corner of the plate (nearest to the MCP-PMTs) are shown in Table~\ref{tab:positions}. These geometrical configurations were chosen such that the Cherenkov patterns were well-contained within the instrumented region of the MCP-PMT detector plane.

\section{Data processing}
\label{sec:dataprocessing}

At the first stage of data analysis, various processing steps were undertaken to calibrate the data. 
All calibrations were data-driven and performed on a large, dedicated set of data,  independent from the datasets used for TORCH  analysis.
These  calibrations included  corrections for effects introduced by the readout electronics: the time-walk of the NINO chip, time alignment of pixels, and the integral non-linearity of the HPTDC chips. All algorithms are described in Refs.~\cite{Brook:2018qdc, Bhasin2020} and summarised below. 
\begin{itemize}
\item {\bf Pixel-to-pixel time alignment and time-walk calibration}: Different routing-line lengths on the NINO and HPTDC boards give small pixel-to-pixel time offsets. These corrections were determined from the requirement that  hits within a photon cluster  have the same arrival time.  Using the same procedure, a  timing slew (time-walk) calibration was made for each pixel, and a correction subsequently made to the arrival time of each photon hit according to the width of its time-over-threshold.    

\item {\bf HPTDC integral non-linearity (INL)}: The HPTDC ASICs which digitise the arrival times and time-over-threshold from the NINO chip employ time-binning which is not equally spaced in time (integral non-linearity).  
A calibration was adopted to correct for the channel-to-channel non-linearity.

\item {\bf Clustering algorithm}: The MCP-PMT is designed such that a single incident photon will induce charge deposits over several neighbouring pixels, in order to improve the effective granularity in the vertical direction. In the present analysis, a simple average position is taken for the centroid of the pixel hits within a  cluster.
Charge-to-TOT calibration of the NINO chip would allow a more accurate cluster centroiding to be performed, and is being developed for future test beams. 

\end{itemize}

\section{Test beam analysis}
\label{sec:analysis}

Analysis of the test beam data 
is categorised into three parts. First, the single-photon time resolution is determined and compared to the target of 70\ps per photon. The contributions  from the intrinsic MCP-PMT response, electronics readout and photon reconstruction are inferred. Studies of the resolution for paths of photons involving side reflections in the radiator are then performed. 
Second, the yields of photons per charged track are determined and compared to simulation. 
The relative yields of photons undergoing higher-order reflections in the radiator are also assessed.
Finally, the per-track timing resolution is measured for tracks yielding multiple photon clusters, 
verifying the improvement in  performance
when clusters  are combined. 
All the above analyses employ data taken at Positions 1, 3, 4, and 5 in the radiator; the exception being for  studies of variation of timing resolution and yield with photon path, where the Position 6 dataset is used.

\begin{figure}[!ht]
    \centering
    \begin{subfigure}[]{0.60\linewidth}
        \includegraphics[width=\linewidth]{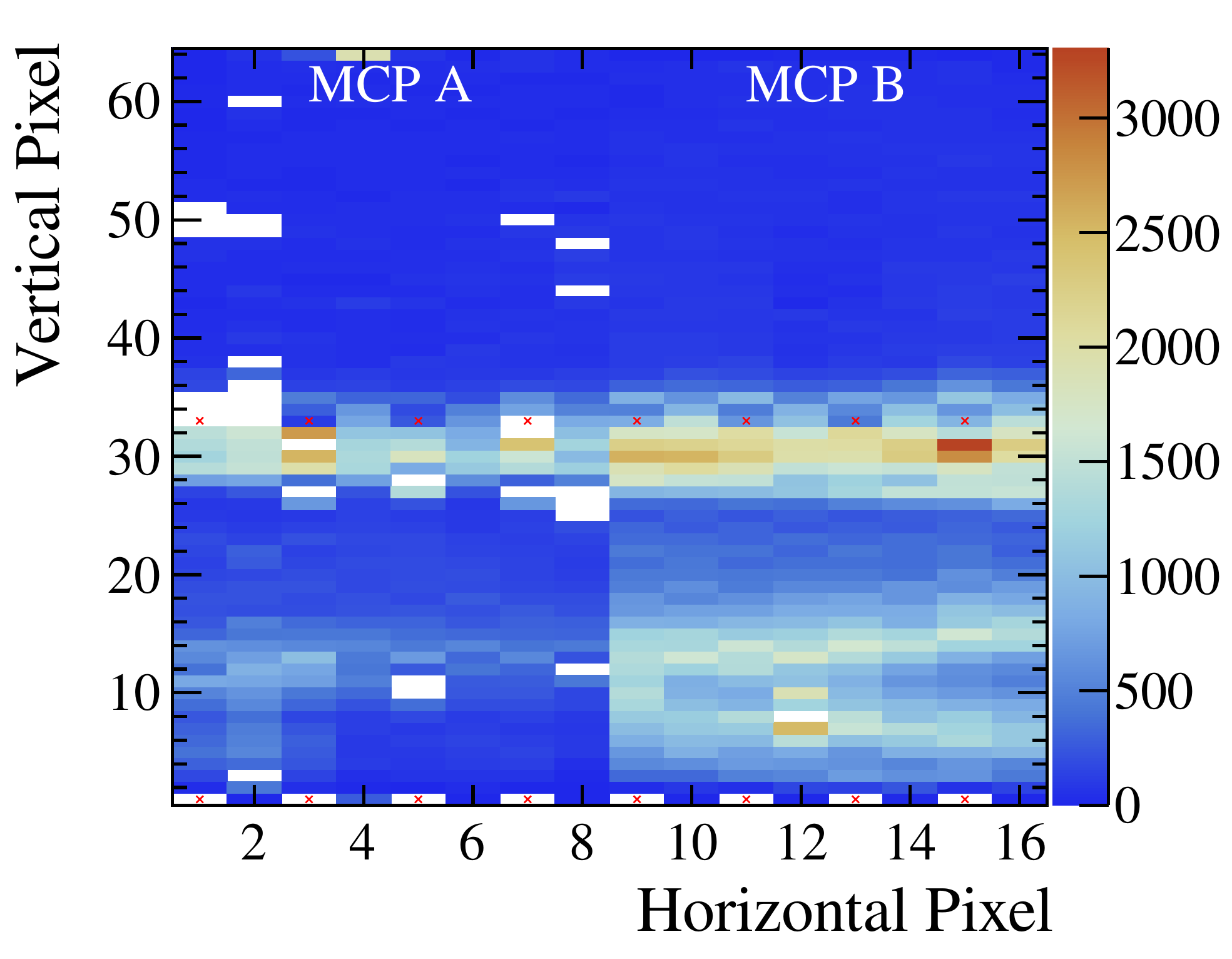} 
        \caption{}
        \label{fig:hitmap}
    \end{subfigure}  \\
        \begin{subfigure}[]{0.75\linewidth}
        \includegraphics[width=\linewidth]{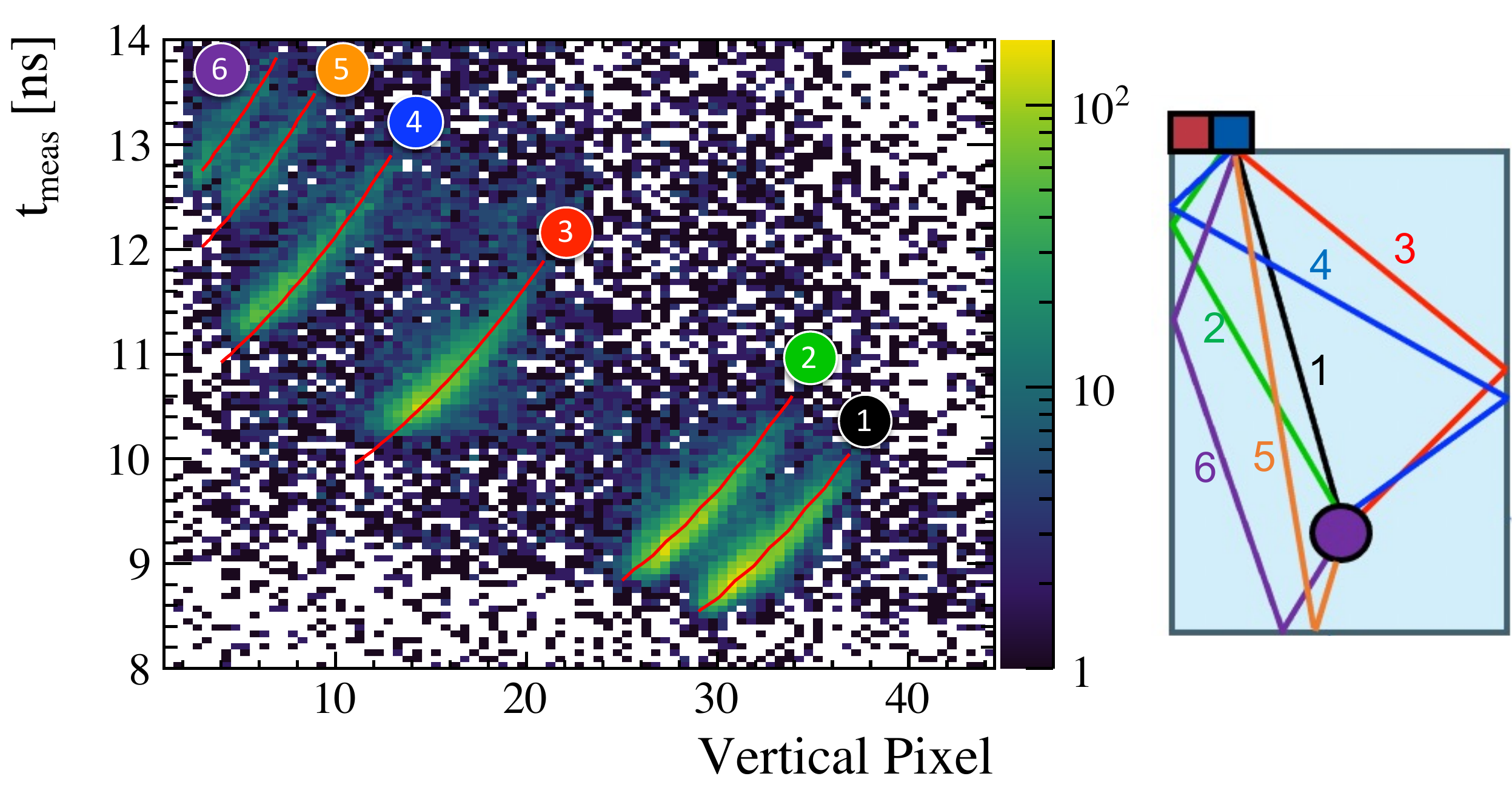}
        \caption{}
        \label{fig:timeprojection}
    \end{subfigure}
    \caption{(a) The spatial distribution of reconstructed photon clusters on MCPs A and B. Pixels marked with a red cross are those where the time reference signals were injected. 
    (b) The time projection of reconstructed photon clusters on MCP-PMT column 16, with the photon trajectories that fall within the MCP-PMT acceptance labelled, and the associated reflections on the right.
    }
    \label{fig:beamtestdata}
\end{figure}

During the beam test,  Proto-TORCH was instrumented with two MCP-PMTs corresponding to 1024 readout channels,  16 of which were allocated to time-reference channels. 
Figure~\ref{fig:hitmap} shows the spatial distribution of reconstructed photon clusters on the MCP-PMT plane for data taken with the beam entering the plate at Position 6. 
The cones of Cherenkov radiation emitted by the charged hadrons are focused into hyperbolic patterns, which are folded in on themselves when the light is reflected from the sides of the module. 
Signals from different paths in the radiator can be detected in the same pixel but will arrive at different times. 
The different QEs of the two MCP-PMTs is responsible for the lower level of light seen in MCP\,A.
The empty (dead) channels, represented by white entries in Fig.~\ref{fig:hitmap}, reduces from MCP\,A to MCP\,B as a result of improvements made in the wire bonding procedure of the NINO chips when mounting on the  boards during manufacture. 

The time-projection plot (Fig.~\ref{fig:timeprojection}) shows the reconstructed photon  arrival time (relative to the reference time) as a function of the clusters' vertical pixel position, for data collected on a typical single MCP-PMT column (horizontal pixel). 
The red lines are predicted arrival patterns, simulated from knowledge of the hadron track position and angle. 
In this projection, the different \textit{orders of reflection} (photon paths in the radiator) are well-separated. The widths of each order 
are then measured to determine the single-photon time resolution.
The  photon yields are determined by defining contained regions around each order in this projection, where the regions are tuned to minimise background counts and to avoid incorrect  counting across neighbouring orders.

\subsection{Single-photon time resolution}
\label{sec:singlephotonres}

The single-photon time resolution is determined using simultaneous unbinned maximum-likelihood fits to the photon arrival times with respect to the time-reference signal ($t_0$). 
A separate fit model is created for each vertical pixel comprising:
\begin{itemize}
    \item Peaking structures to fit the signal distributions, modelled by Crystal Ball functions \cite{Skwarnicki:1986xj} which fit asymmetric Gaussian distributions. This function models the tails caused by late arriving back-scattered photoelectrons from the MCP surface. Analytical reconstruction predictions are used to determine how many peaking structures are expected for each pixel;
    \item Peaking background shapes to account for structures observed in the data, modelled by Gaussians. The structures are bands across all pixels that appear at the same time, attributed to photons from secondary sources and/or scattered signal light;
    \item Polynomial background distributions to account for observed random background levels. 
\end{itemize}

\begin{figure}[t]
    \centering
    \begin{subfigure}[]{0.50\linewidth} 
        \includegraphics[width=\linewidth]{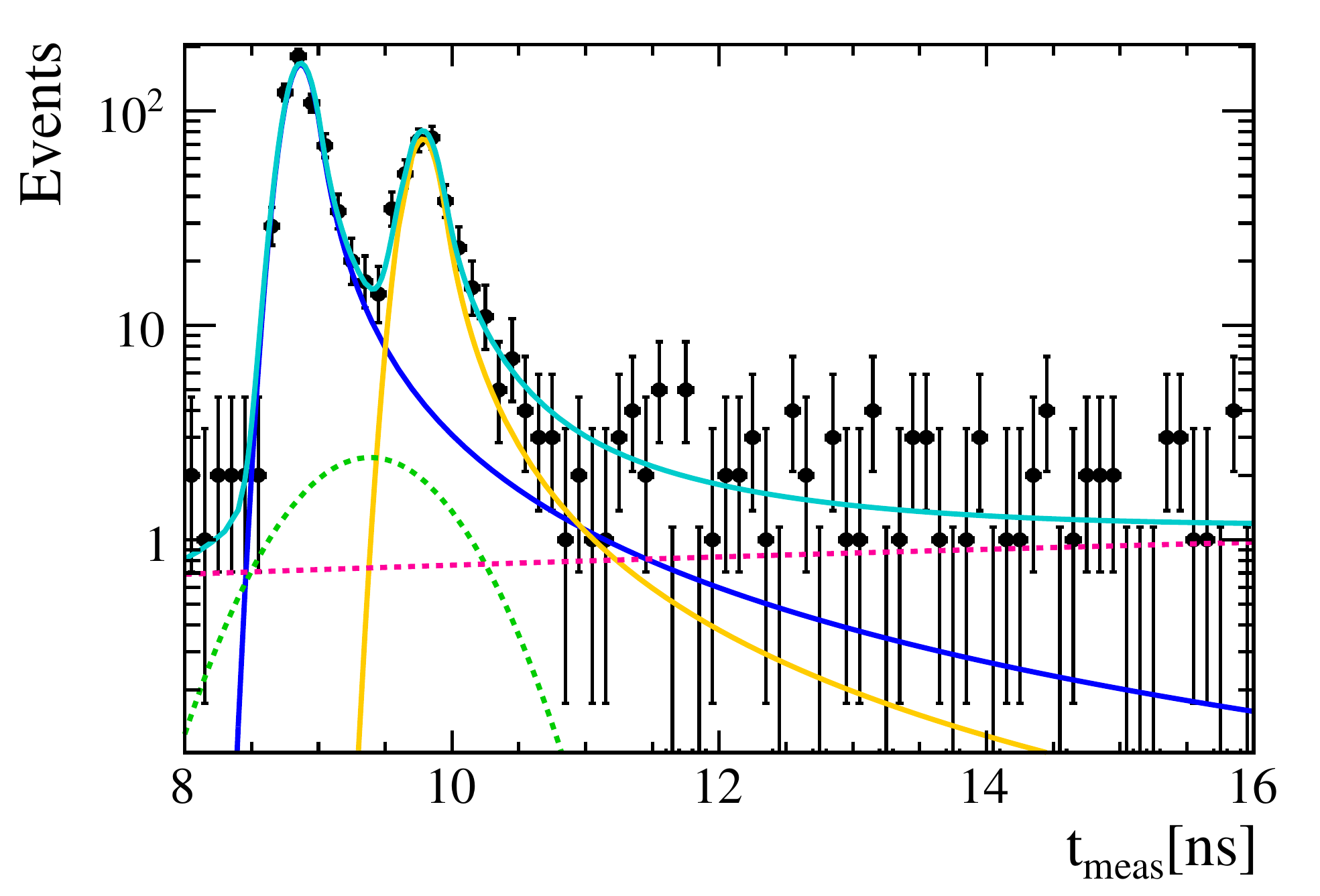}
        \caption{}
        \label{fig:time_resolution_fit}
    \end{subfigure}
    \begin{subfigure}[]{0.49\linewidth}
        \includegraphics[width=\linewidth]{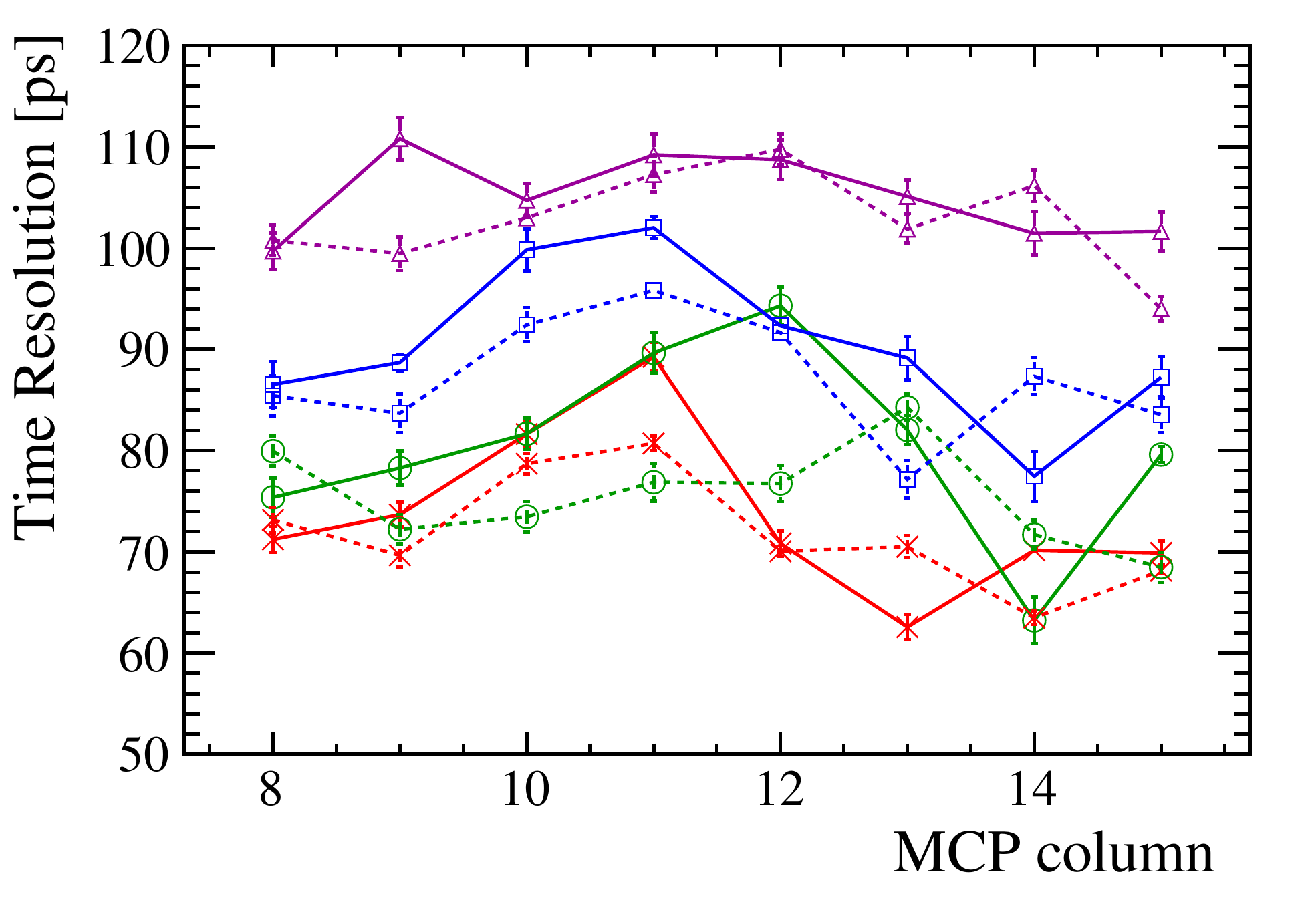}
        \caption{}
        \label{fig:time_resolution_percol}
    \end{subfigure}
    \caption{
    (a) An example of a fit to the time of arrival of photons on a single vertical pixel (pixel 30 as shown in Fig.~\ref{fig:timeprojection}), used to measure the single photon time resolution. On this pixel there are two orders of reflection, with times of arrival that peak at 8.9\,ns and 9.8\,ns. Their respective measured widths are $124\pm3$\ps and $134\pm3$\ps. 
    In addition to the two signal peaks (blue / yellow), the background contributions (dotted) comprise the total fit (cyan).
    (b)
    The single-photon time resolution determined in different columns of MCP\,B and for different beam positions (Positions 1 ($\times$), 3 ({\Large$\circ$}), 4 ($\rlap{$\sqcap$}\sqcup$), 5 ($\triangle$) as indicated in Fig.~\ref{fig:torchpos}).
    The full (dotted) lines are the results from the pion (proton) samples.
    }
    \label{fig:time_resolution}
\end{figure}

A simultaneous fit to the photon time distribution is performed across all vertical pixels of a column.
In each vertical pixel the yields, mean times and widths of the Crystal Ball probability distribution functions (PDFs) are free to float, however the widths for individual reflection orders are shared in the fit across all vertical pixels in the same MCP-PMT column. The parameters of the background PDFs are also shared across all vertical pixels in a column.
The width of the Gaussian component of the fitted Crystal Ball function gives the measured resolution per order of reflection per column, $\sigma_\text{meas}$. 
An example fit to a single pixel is shown in Fig.~\ref{fig:time_resolution_fit}  in which two peaking orders of reflection are observed, with  one peaking background structure and the polynomial background. 

The measured resolution $\sigma_\text{meas}$  receives contributions from the experimental setup that are external to TORCH, which need to be subtracted. 
Here the start-time resolution  is the primary contribution. During nominal data taking, the rate of hits is maximised by requiring only the downstream timing-reference scintillator stations to have fired. From a study of the time difference between the two timing signals for different trigger configurations, the $t_0$ resolution, $\sigma_{\text{time ref}}$, is measured to be $53.2\pm 0.6\ps$. 
The second contribution is due to the finite size of the beam; the uncertainty in the  particle entry position in the quartz plate results in a variation of the photon path length and hence its arrival time.
From simulation and a knowledge of the beam profile, the contribution to the time resolution, denoted by $\sigma_{\text{beam}}$, is between 15.1 and 30.4\ps~\cite{Hancock:thesis}, dependent on the height at which the beam enters the  plate.
The above contributions are subtracted in quadrature from $\sigma_\text{meas}$ to obtain the TORCH single photon time resolution:

\begin{equation} \label{eqn:res_contibutions}
    \sigma_{\text{TORCH}}^{2} = \sigma_{\text{meas}}^{2} - \sigma_{\text{time ref}}^{2} - \sigma_{\text{beam}}^{2}. 
\end{equation}


The time resolution is measured for several beam entry positions, indicated by Positions 1, 3, 4 and 5 in Fig.~\ref{fig:torchpos}. Here the distance from the top of the plate to the beam is varied from 175\mm to 1115.4\mm in four steps. 
The TORCH time resolution for each MCP-PMT column and beam position is shown in Fig.~\ref{fig:time_resolution_percol}, separately for the samples of pions and protons. 
It can be seen that the time resolution of the prototype approaches the design goal of 70\,ps but there is some degradation for entry positions far from the MCP-PMTs, which is to be expected.
Due to the finite size of the pixels, the uncertainty on the position and the Cherenkov angle introduces an uncertainty in the measurement of the photon energy, hence the time resolution. This uncertainty scales linearly with path length and propagation time. 
As a compensation for this degradation, in any future PID reconstruction, there will be  additional information from the time-of-propagation  measurement  for the larger photon path lengths, where the difference in Cherenkov angle  of any two particle species will result in a spatial and  timing separation at the photodetector plane. 

\begin{figure}[t]
    \centering
    \includegraphics[width=0.7\linewidth]{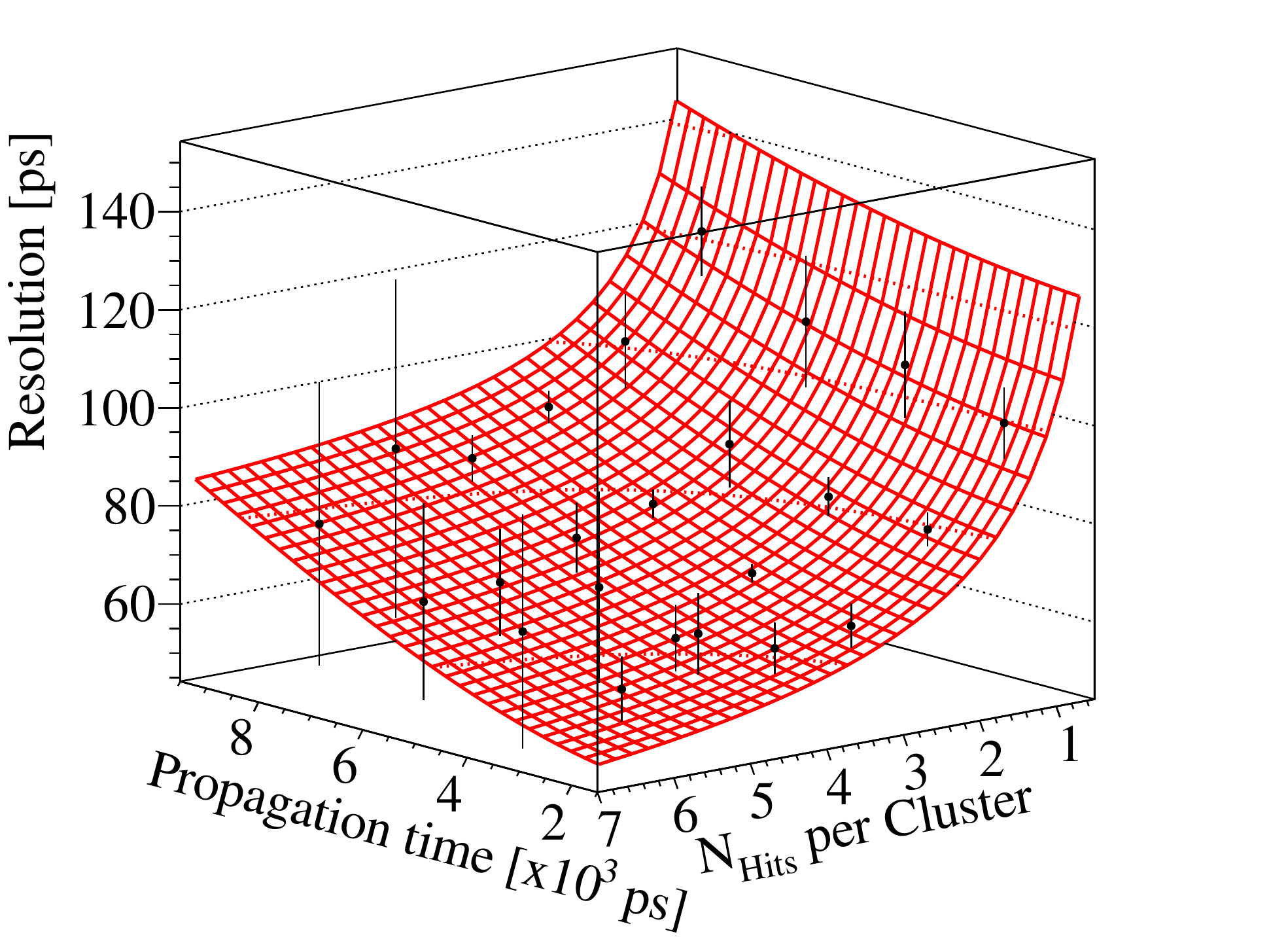}
    \caption{The TORCH time resolution of the proton sample, displayed as a function of the propagation time and number of hits per photon cluster. The overlaid surface is the result of the 2D fit described in the text. }
    \label{fig:time_resolution_2D}
\end{figure}

To quantify the contributions to the TORCH time resolution from different sources, a two-dimensional fit is performed as a function of the photon propagation time (directly related to the beam entry position) and photon-cluster size, shown for protons in Fig.~\ref{fig:time_resolution_2D}. The time resolution is parametrised by 

\begin{equation} \label{eqn:res_2d}
    \sigma_{\text{TORCH}}^{2} = \sigma_{\text{prop}}(t_{p})^{2} + \sigma_{\text{MCP}}^{2} + \sigma_{\text{RO}}(N_{\text{Hits}})^{2},
\end{equation}
where the term $\sigma_{\text{prop}}(t_{p})$   characterises the contribution  that grows with photon propagation time, $t_{p}$;  a
constant term $\sigma_{\text{MCP}}$ is associated to the resolution of the MCP-PMT;    
a term $\sigma_{\text{RO}}(N_{\text{Hits}})$ describes effects from the charge-spread and readout process which is a function of the number of hits in a cluster, $N_{\text{Hits}}$.  A linear dependence on $t_{p}$ and a $1/\sqrt{N_\text{Hits}}$ dependence  on ${N_\text{Hits}}$ are assumed in $\sigma_\text{prop}$ and $\sigma_\text{RO}$, respectively.
The results of the fits are presented in Table \ref{tab:2dfitres}, together with target values, which assume a 33\,ps intrinsic MCP-PMT resolution added in quadrature with a $60\,\text{ps}/\sqrt{N_\text{Hits}}$ readout resolution, giving 50\,ps for 2.5 hits per cluster. The time-of-propagation resolution assumes a 50\,ps resolution based on a photon propagation time for a 2.5m quartz length. 
Whilst the data do not entirely match expectations for the running of the $\sigma_\text{prop}$ and $\sigma_\text{RO}$ terms, the MCP resolution term is in agreement. 
The latter two resolutions are expected to improve with future  electronics calibrations, and the sources of these contributions will be further studied in laboratory tests. 

\begingroup
\renewcommand{\arraystretch}{1.3} 
\begin{table}[t]
\caption{The results of a 2D fit to the single photon time resolution as a function of propagation time, $t_p$ in \ps, and the number of hits per photon cluster. The results of analysis of both the pion and proton samples are shown separately and are consistent with each other. The target performance is also shown.
}
\begin{tabular}{ |c|c|c|c| } 
\hline
\multirow{2}{*}{Contribution} & \multicolumn{2}{|c|}{Fitted values (ps)}                                                & \multirow{2}{*}{Target values (ps)}                                           \\ \cline{2-3}
                                & Pion                                          & Proton                                  &  \\ \hline 

$\sigma_{\text{prop}}(t_p)$     & $(8.3\pm0.7)\times 10^{-3} \times t_{p}  $       & $(7.6\pm0.5)\times 10^{-3}\times t_{p} $  & $(3.75\pm0.8) \times 10^{-3}\times t_{p} $ \\ 
$\sigma_{\text{MCP}}$         & $34.5\pm8.6$                                    & $31.0\pm7.6$                               & $33$  \\ 
$\sigma_{\text{RO}}(N_{\text{Hits}})$  & $(96.2\pm6.7) / \sqrt{N_{\text{Hits}}}$   & $(95.0\pm6.0)/ \sqrt{N_{\text{Hits}}}$ &  $60 / \sqrt{N_{\text{Hits}}}$\\ 
\hline
\end{tabular}
\label{tab:2dfitres}
\end{table}
\endgroup

Data taken with the beam entering the plate at Position 6, a location which is centrally located towards the bottom of the plate,  are used to study the time resolutions of photons  that have undergone multiple side and/or bottom reflections in the plate, as shown in Fig.~\ref{fig:timeprojection}. 
Photons collected on MCP\,B are studied, where there is good separation between different orders of reflection across all columns.

\begin{figure}[ht]
    \centering
    \begin{subfigure}[]{0.49\linewidth}
        \includegraphics[width=\linewidth]{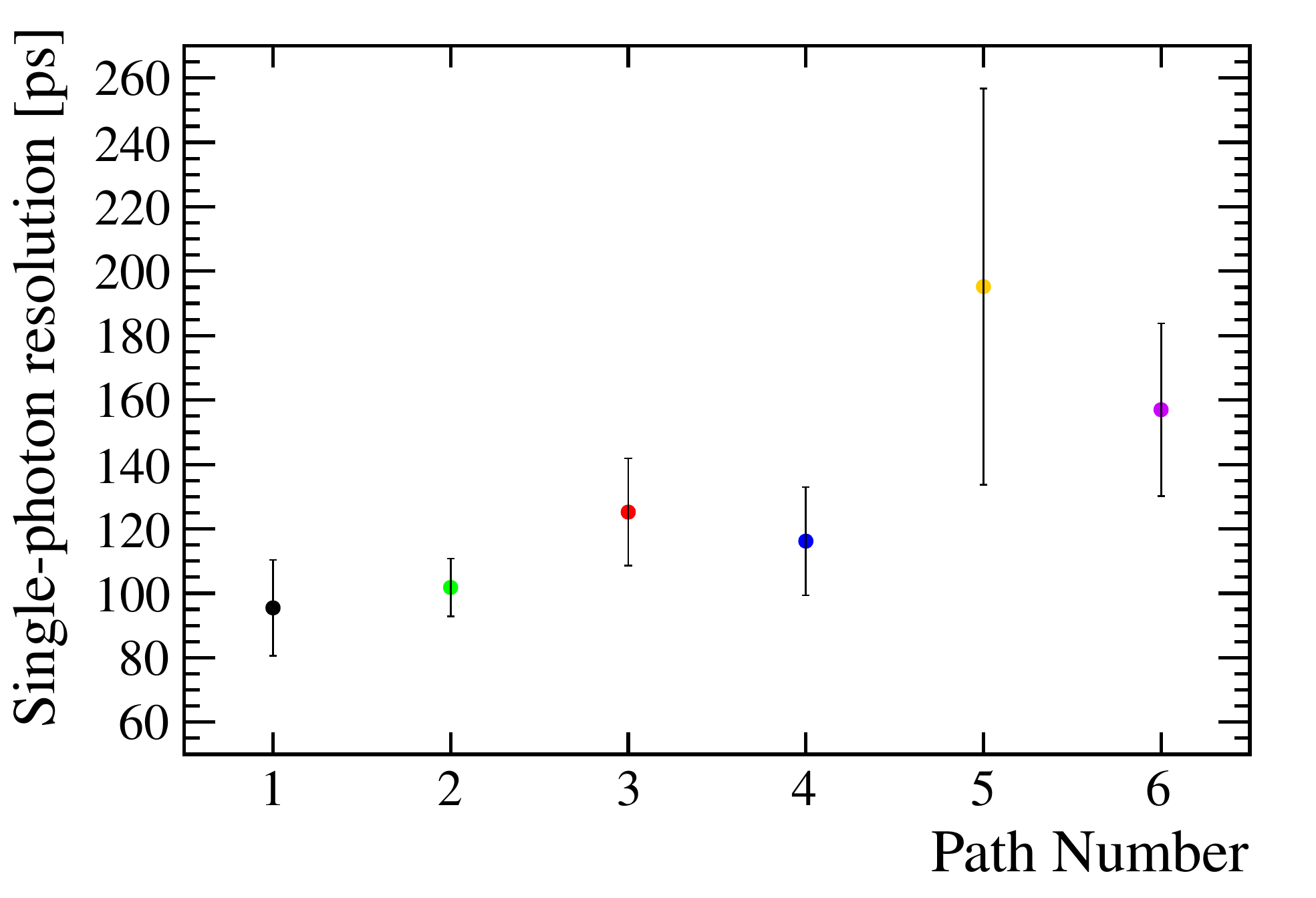} 
        \caption{}
        \label{fig:Pos6_singleTR_bypath}
    \end{subfigure}
    \centering
    \begin{subfigure}[]{0.49\linewidth}
        \includegraphics[width=\linewidth]{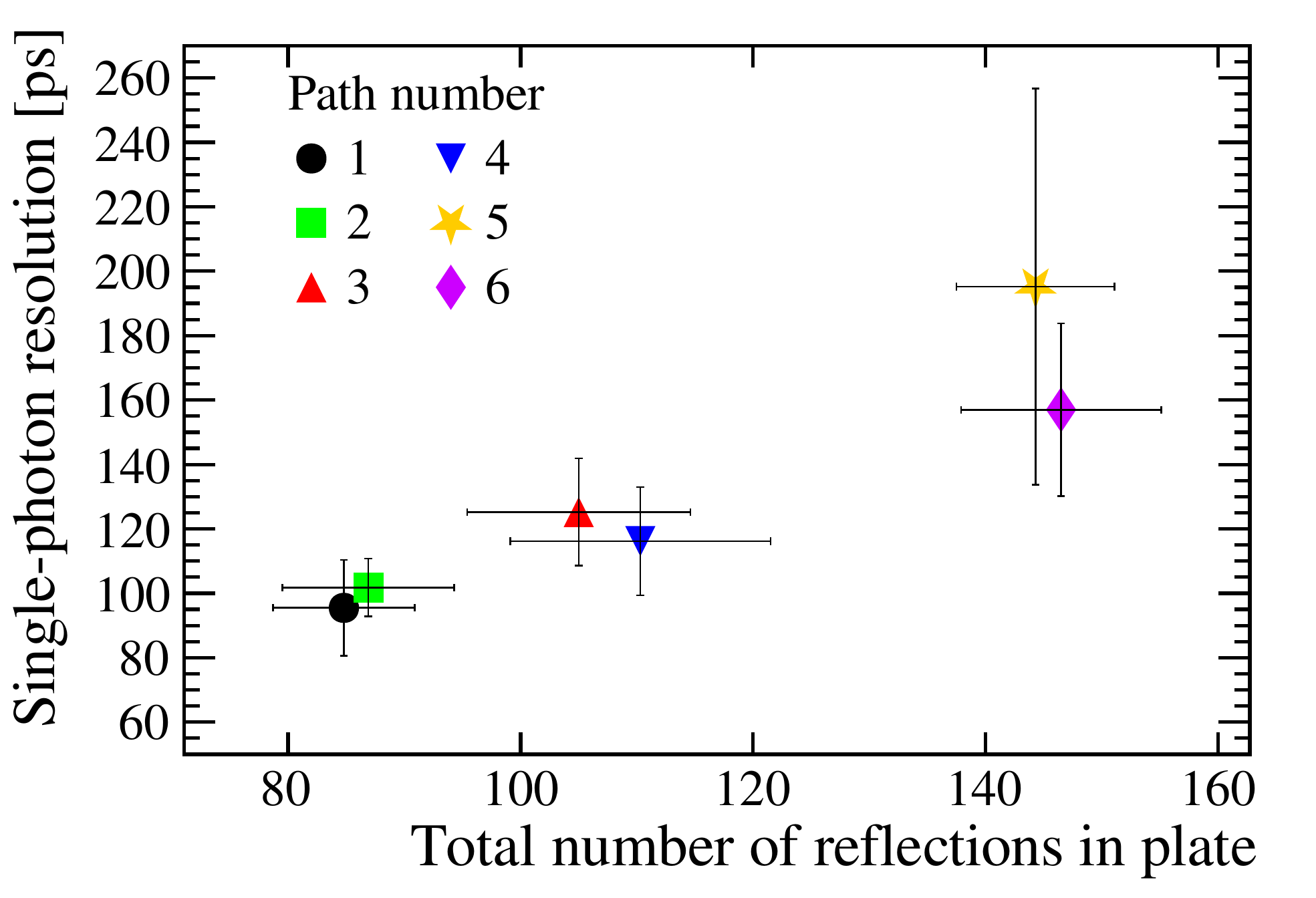} 
        \caption{}
        \label{fig:Pos6_singleTR_byrefl}
    \end{subfigure}
    \caption{(a) Single-photon time resolution for different photon paths originating from pions entering the plate at Position 6.
    (b) Single-photon time resolution as a function of the total number of reflections made inside the quartz radiator (taken from simulation) for different photon paths. The colours indicate the paths as shown in Fig.~\ref{fig:timeprojection}. 
    }
    \label{fig:time_resolution_pos6}
\end{figure}

The simultaneous fit, as described above, is used to determine the resolutions of the six orders of reflection in each column. 
For each order, a weighted average of the measured resolutions across all columns is calculated and is shown in Fig.~\ref{fig:Pos6_singleTR_bypath}. 
Figure~\ref{fig:Pos6_singleTR_byrefl} shows the resolution as a function of the total number of photon reflections in the quartz radiator, determined from simulation. The  number of reflections from the front and back surfaces  as well as those from the side and/or bottom faces are counted, although   dominated by reflections from the front/back. The vertical error bars indicate the systematic uncertainties on the measurements associated with the column averaging, whilst the horizontal error bars indicate the range of the number of
reflections.

The measured resolution for the direct photon path (path number 1) is approximately 100\,ps which is similar to the measured resolution of direct photons from Position 5. Photon paths with side and bottom   reflections have slightly  degraded resolutions, to be expected from the increased path length and total numbers of reflections.

\subsection{Photon yields}
\label{sec:photoncounting}
 
The number of photons collected in Proto-TORCH are counted and compared to yields obtained from simulated samples. 
To exclude photons that are likely to have originated from background sources, photons are only counted if they arrive within $0.5\ns$ of the peak time of arrival, and on pixels close to expectation (determined from simulation). 

\begin{figure}[t]
    \centering
    \begin{subfigure}[]{0.45\linewidth}
        \includegraphics[width=\linewidth]{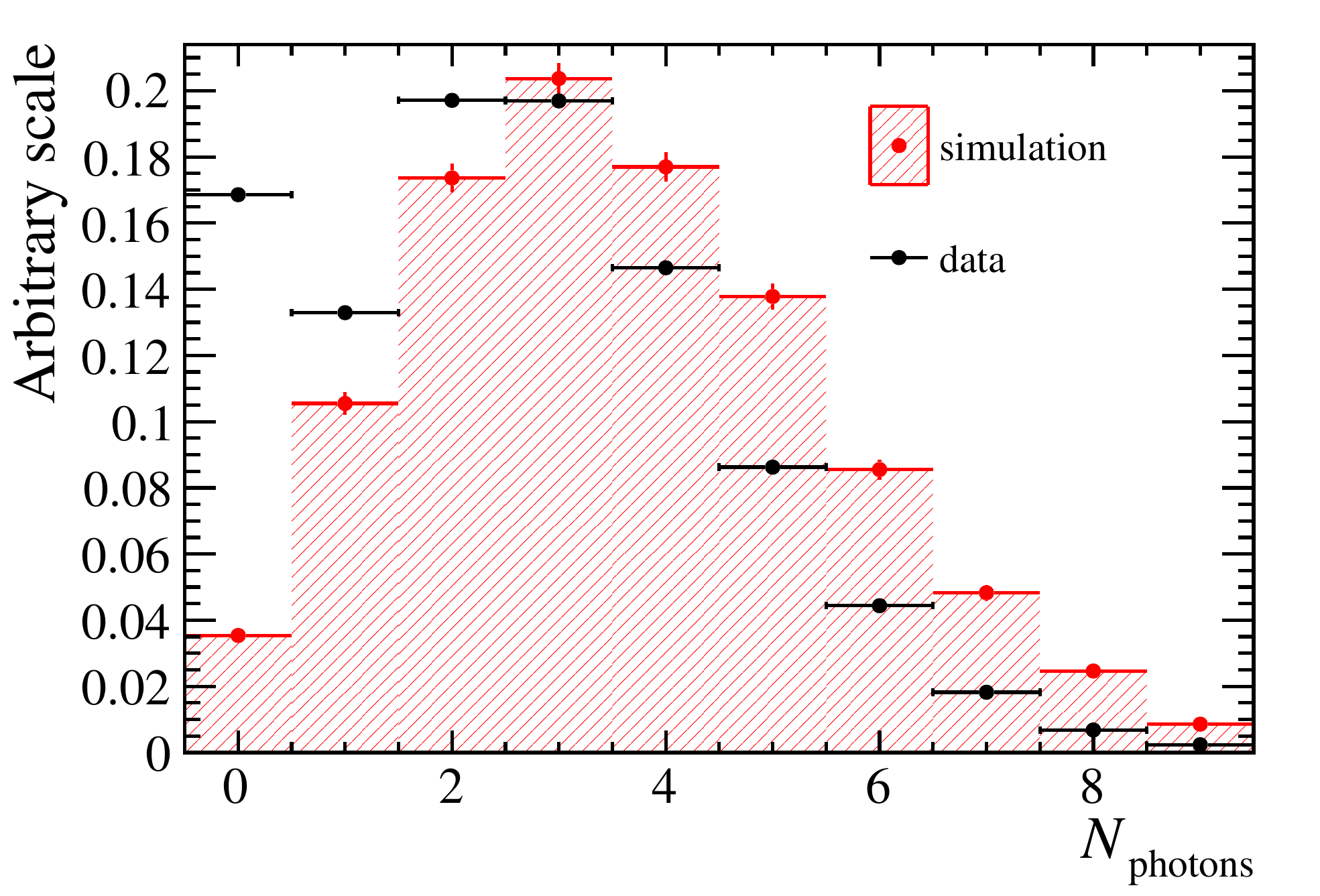} 
        \caption{Position 1}
    \end{subfigure}
    \begin{subfigure}[]{0.45\linewidth}
        \includegraphics[width=\linewidth]{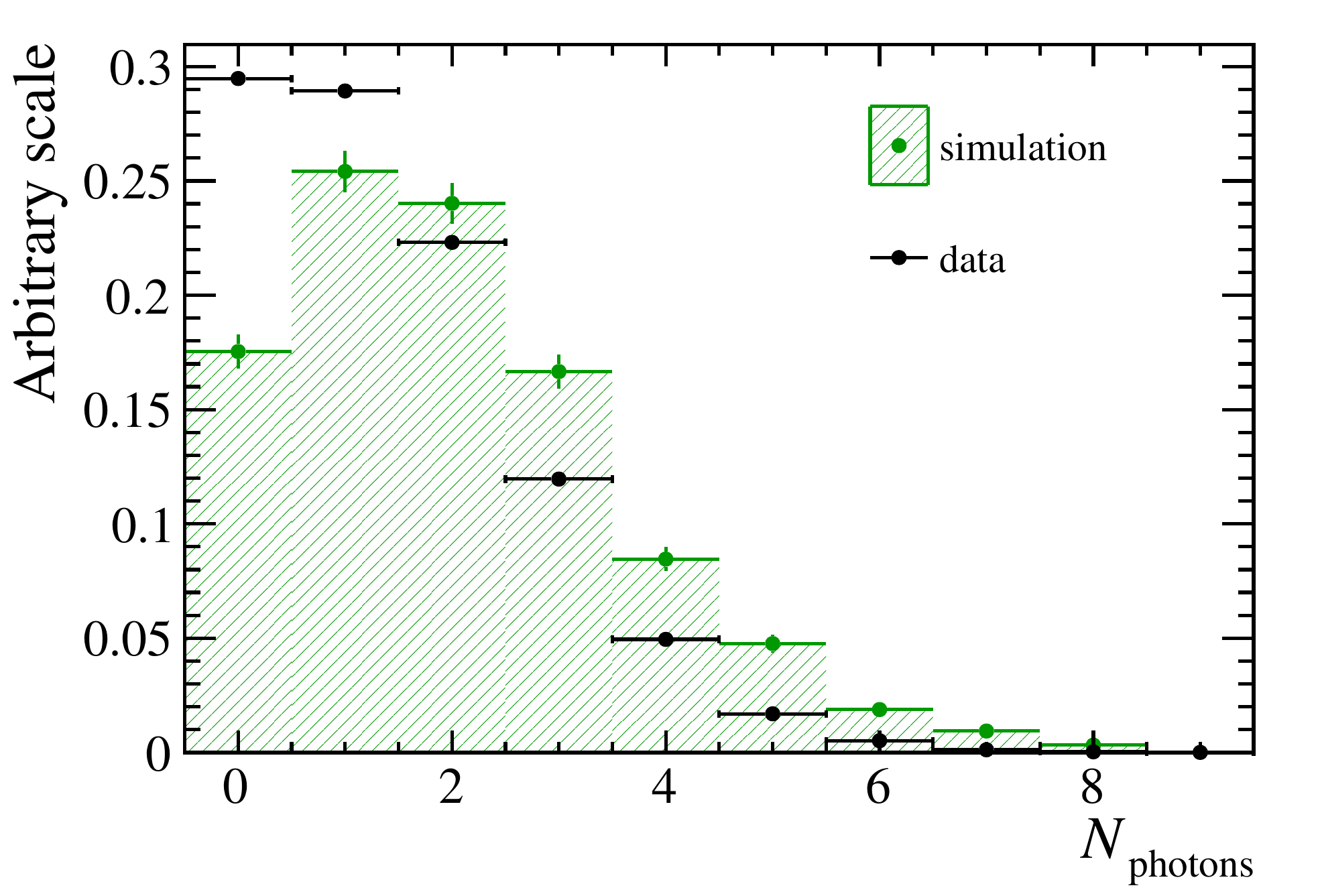}
        \caption{Position 3}
    \end{subfigure}
    \begin{subfigure}[]{0.45\linewidth}
        \includegraphics[width=\linewidth]{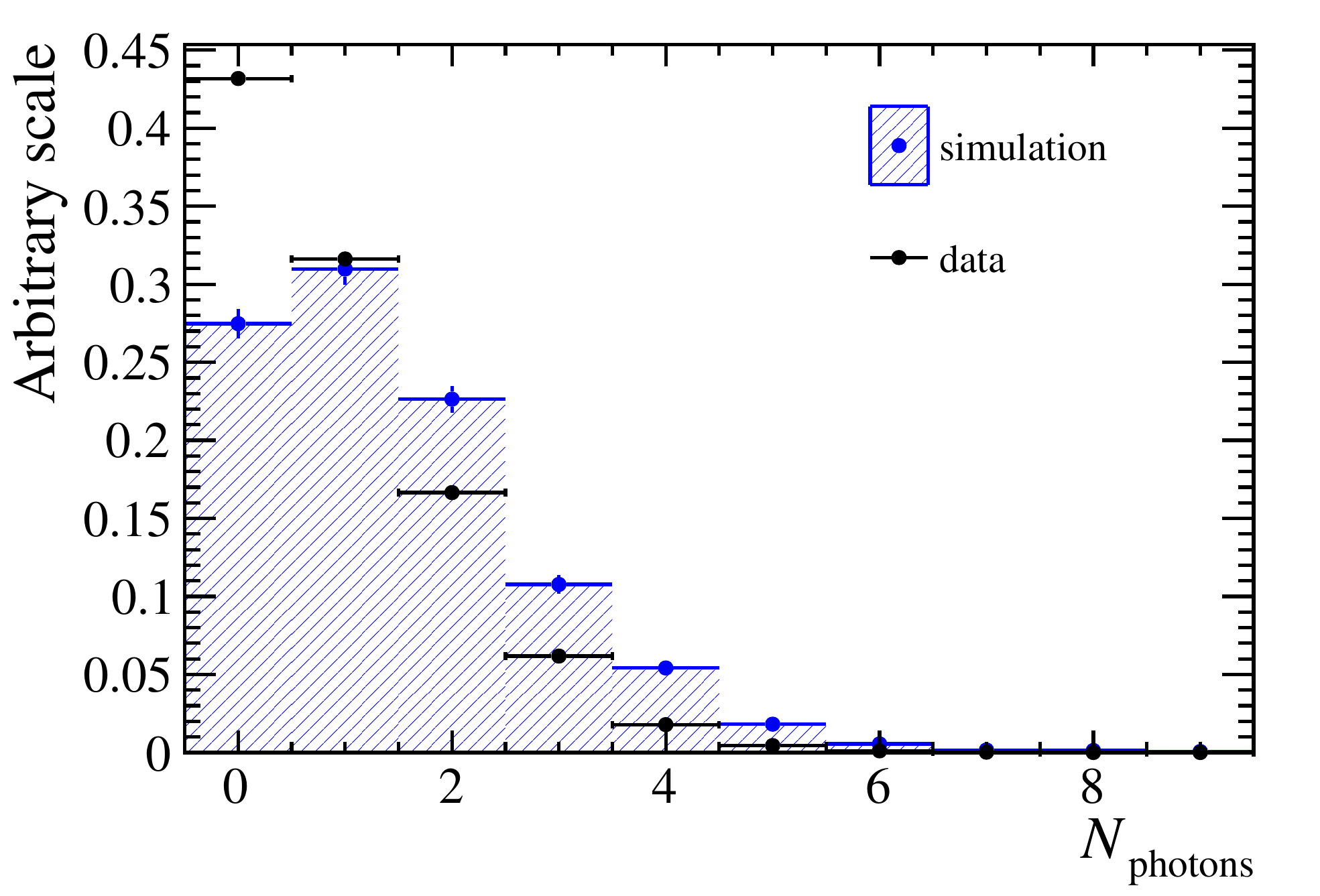}
        \caption{Position 4}
    \end{subfigure}
    \begin{subfigure}[]{0.45\linewidth}
        \includegraphics[width=\linewidth]{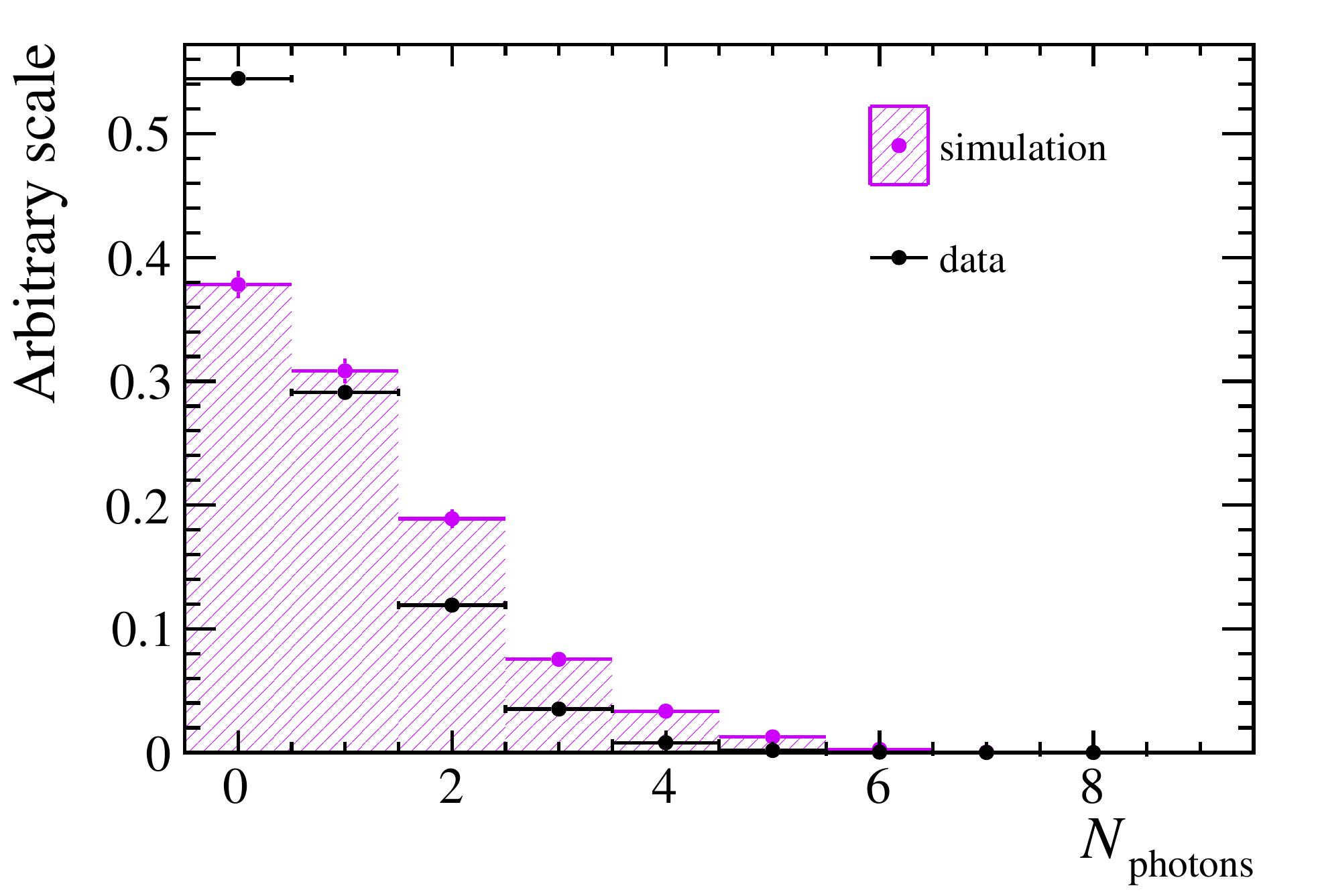}
        \caption{Position 5}
    \end{subfigure}
    \caption{Photon counting yields for the four beam entry positions from data and simulation.}
    \label{fig:photon_counting}
\end{figure}

For beam positions close to the side edge of the quartz, only photons that travel directly or are reflected once off the nearest side edge are counted.
The yields of the number of photons detected per incident hadron in data, $N_\text{photons}$,  are compared to expectations from simulation in Fig.~\ref{fig:photon_counting}. 
Some discrepancies between data and simulation are observed; in particular there are more events in data where no photons are observed   than in simulation. 
This is attributed to two effects: firstly the likely occurrence of a small number of false triggers, and secondly that  small charge-shared pulses can have amplitudes below the NINO thresholds which are not accurately modelled in the simulation. 
The mean number of photons detected per hadron for each beam position is shown in Table \ref{tab:photon_counting}. 
Improved and consistent agreement is seen when events with  $N_\text{photons}=0$ are excluded, better than 80\% at all beam positions. The photon yield reduces when the beam is positioned further from the MCP-PMTs because the angular acceptance of the detection surface decreases.
Note that the mean photon yield per hadron is significantly reduced compared to future expectations of 30 photons per charged track for a fully-instrumented module. 
This is primarily because only two (of the eleven) MCP-PMTs are instrumented, and only two of the possible photon paths are considered here. In addition, future MCP-PMTs are expected to have a much improved quantum efficiency which will result in larger yields. 

\begin{table}[t]
\caption{The mean number of photons detected originating from Positions 1, 3, 4 and 5 in data and simulation. The final column gives the ratio of the mean yields when the contributions from events with zero photons detected in data and simulation are excluded.}
\begin{tabular}{ |c|c|c|c|c| } 
\hline
\multirow{2}{*}{Position} & \multicolumn{2}{|c|}{Mean $N_{\text{photons}}$} & \multicolumn{2}{|c|}{Mean(Data)/Mean(Simulation)} \\ \cline{2-5} 
& Data & Simulation & All & Excluding $N_{\text{photons}}$=0\\ \hline 
1 & 2.605 $\pm$ 0.007 & 3.586 $\pm$ 0.020 & 0.726 $\pm$ 0.004 & 0.843 $\pm$ 0.005 \\ 
3 & 1.419 $\pm$ 0.005 & 2.016 $\pm$ 0.029 & 0.704 $\pm$ 0.010 & 0.824 $\pm$ 0.010 \\
4 & 0.937 $\pm$ 0.004 & 1.454 $\pm$ 0.024 & 0.644 $\pm$ 0.011 & 0.823 $\pm$ 0.008 \\
5 & 0.677 $\pm$ 0.002 & 1.127 $\pm$ 0.022 & 0.600 $\pm$ 0.012 & 0.820 $\pm$ 0.009 \\
\hline
\end{tabular}
\label{tab:photon_counting}
\end{table}


The yields of photons that have undergone multiple side reflections within the quartz plate are also studied, using data taken with the beam at Position 6. This central location  towards  the base of the plate allows for reflections from both side surfaces and the bottom surface of the radiator to be observed. 
As shown in Fig.~\ref{fig:timeprojection}, some orders of reflection arrive on the same pixel close in time. To minimise double counting and mis-association, the boundaries around each order are adapted from the nominal $\pm 0.5\ns$ region for each column individually. 
The yields from different photon paths are presented in Table \ref{tab:pos6}  and  good agreement  is generally seen in the photon yields when compared to simulation.
The deficiency in photon yields in data for photons reflecting from the bottom surface is not completely understood, although we note these photons undergo a significantly higher number of reflections compared to the other paths within the radiator. Possible causes may be a reduced-quality finish of this surface (which is not mirrored) and interference with a mounting support bracket which is in close proximity.

\begin{table}[t]
\caption{The yields of different photon paths originating from Position 6 in data and simulation. The path number corresponds to the paths shown in Fig.~\ref{fig:timeprojection}. The percentage is the fraction of the sum of yields from all paths.
}
\begin{tabular}{ |c|c|c|c|c| } 
\hline
\multirow{2}{*}{Reflections (path number)} & \multicolumn{2}{|c|}{Data} & \multicolumn{2}{|c|}{Simulation}  \\ \cline{2-5}
                    & Yield & \%    & Yield & \% \\ \hline 
Direct (1)          & 47533 & 27    & 23790 & 25 \\ 
One side (left) (2)   & 46300 & 26    & 21422 & 23 \\
One side (right) (3)  & 37383 & 21    & 16458 & 18 \\ 
Two side (4)          & 28936 & 16    & 13957 & 15 \\
One bottom (5)        & 9193  &  5    & 10624 & 11 \\ 
One bottom, one side (6)& 9100  &  5    & 7574 & 8 \\ 
\hline
\end{tabular}
\label{tab:pos6}
\end{table}

\subsection{Per-track time resolution}
\label{sec:tracktimeres}

The final stage of the analysis involves measuring the so-called ``per-track'' timing resolution,  where the best estimate of  the track time is determined from a combination of arrival times of all photon clusters in the event. 
The overall TORCH timing performance relies on there being a $1/\sqrt{N_\text{photons}}$ dependence, assuming   that the measurement properties of individual photon hits are uncorrelated with others from the same track. 
In what follows, the time-stamps of the individual photons from a single incident track as identified in Sec.~\ref{sec:singlephotonres}  are combined,  and the corresponding time resolution determined.

The analysis combines information from photons that arrive on different MCP columns. In order to increase the yield of events that have several detected photons, hits on both MCPs A and B are  used. 
The per-track time resolution is measured as a function of track entry position in the quartz radiator, and as a function of $N_\text{photons}$.  

The time of arrival of each photon, $t_{\text{meas}}$, is compared to the expected time of arrival, $t_{\text{exp}}$, obtained from the peak position of the relevant Crystal Ball function in the single photon analysis.
A variable $\Delta t$ is introduced to provide the best estimate of the average time difference $<t_\text{exp} - t_\text{meas}>$ 
of all photons that originate from the same hadron. The spread in $\Delta t$ then indicates the best estimate of the per-track time resolution. 
The value of $\Delta t$ for each  event is found by constructing a $\chi^2$ variable, 

\begin{equation} \label{eqn:chi2}
    \chi^2 = \sum\limits_{i}^{N_\text{photons}} \left( \frac{t^i_\text{meas} + \Delta t -t^i_\text{exp}}{\sigma^i_\text{meas}} \right)^2, 
\end{equation}
minimisation of 
which with respect to $\Delta t$ 
allows the most probable value of $\Delta t$, 
which is equivalent to the weighted mean.

\begin{figure}[t]
    \centering
    \begin{subfigure}[]{0.45\linewidth}
        \includegraphics[width=\linewidth]{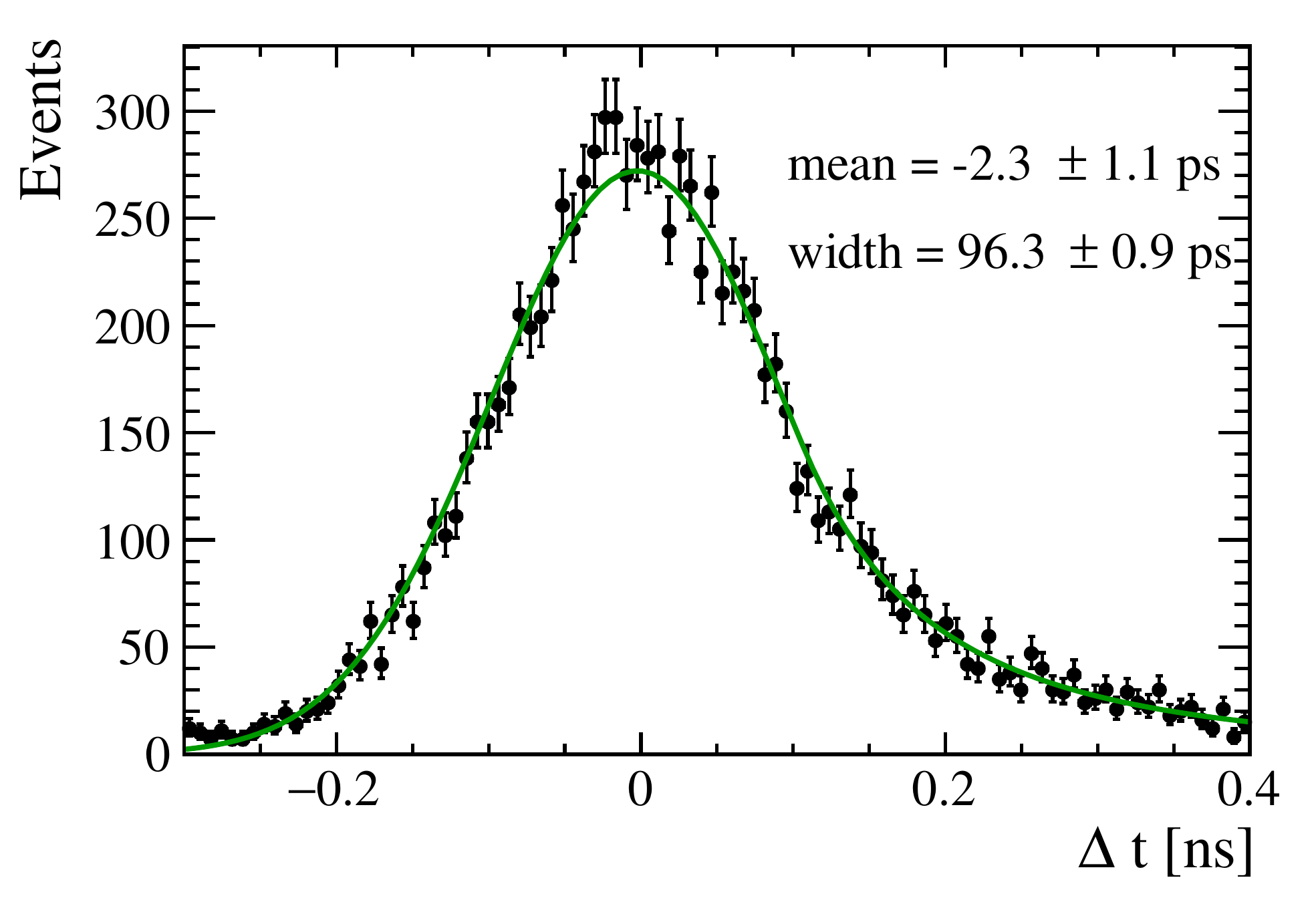}
        \caption{}
    \end{subfigure}
    \begin{subfigure}[]{0.45\linewidth}
        \includegraphics[width=\linewidth]{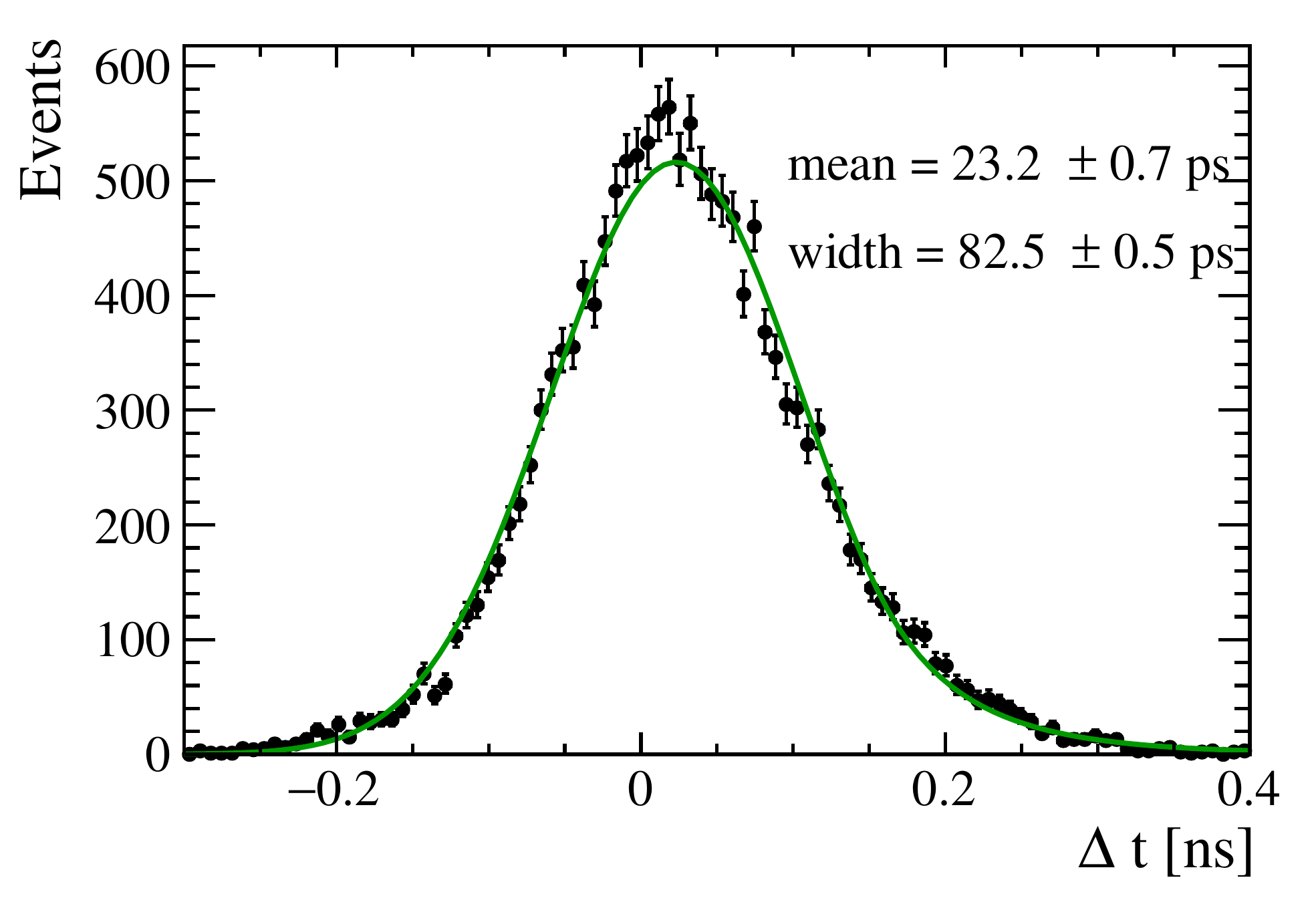}
        \caption{}
    \end{subfigure}
    \caption{The distribution of $\Delta t$ for events with beam entry at Position 1, with (a) one photon detected per event and (b) three. The means and widths specified refer to the means and sigmas respectively of the primary Gaussians in the Crystal-Ball fits. The width quoted is not yet corrected for external time resolution sources. }
    \label{fig:deltat}
\end{figure}

The pion data (with direct photons paths only) are divided into separate samples by the numbers of photons detected per event ($N_\text{photons}$), and by track entry point. 
For each sample, the distribution of $\Delta t$ is fit with a Crystal Ball function to characterise the signal.  
Examples of the $\Delta t$ distribution and fitted PDFs for data taken at Position 1 are shown in Fig.~\ref{fig:deltat}. 
The width of the Gaussian component of the Crystal Ball function is taken as the measured per-track time resolution, which is then corrected by subtracting contributions from the beam spread and time reference according to Eqn.~\eqref{eqn:res_contibutions}. 

\begin{figure}[t]
    \centering
    \includegraphics[width=0.6\linewidth]{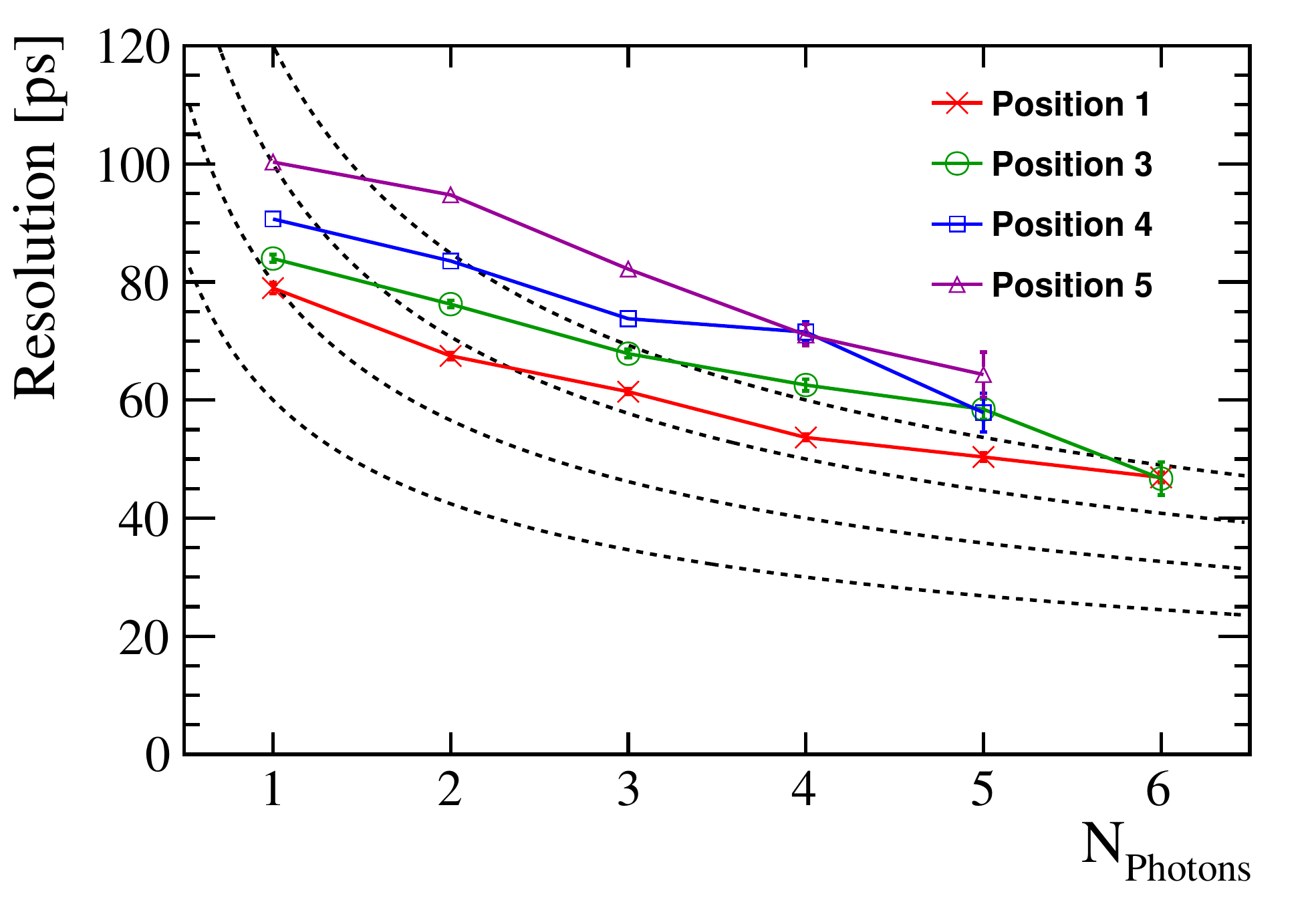} 
    \caption{Proto-TORCH time resolution as a function of the number of collected photons. 
    The dotted lines indicate the function 
    $\sigma_t / \sqrt{N_\text{photons}}$, where $\sigma_t$ = 60, 80, 100, 120\,ps from the bottom upwards.}
    \label{fig:track_time_resolution}
\end{figure}

The per-track time resolution as a function of the number of photons per event is shown in Fig.~\ref{fig:track_time_resolution} for different beam entry positions. 
The measured resolution decreases as the number of photons increases, but more slowly than a $1/\sqrt{N_\text{photons}}$ dependence. The 1/$\sqrt{N_\text{photons}}$ dependence assumes that the underlying resolution is the same for all photons however the greater effect of backscattered tails for low $N_\text{photons}$, seen in Fig.~\ref{fig:deltat}, is a likely cause of a deviation from the expected behaviour. A further systematic effect can be associated with the clustering procedure, common across all hits. In addition, correlations are possible within the electronics between closely-spaced multiple photon hits: a common clock which defines the 100\,ps TDC binning services 64 adjacent channels of the HPTDC chips, which could contribute to deviations from an uncorrelated  100/$\sqrt{12}$\,ps bin resolution. A significantly finer time bin designed into the next generation of TDCs will alleviate this issue. 
Such features still need to be fully understood so that improvements in the resolution can be made in the future.

\section{Conclusions}
\label{sec:conclusions}

Results have been reported from a half-height, full-width LHCb TORCH prototype module, using samples of 8\gevc pions and protons from the CERN T9 test beam. 
The prototype was instrumented with two custom-built MCP-PMTs (out of a possible eleven). 
Photon yields from different track locations are consistent with the expectations from simulation, indicating that acceptance effects and sources of efficiency are largely understood. 
Single-photon time resolutions of the prototype have been measured for different beam entry positions in the quartz radiator plate. 
For entry positions close to the MCP-PMTs, the desired 70\,ps time resolution is achieved.
For positions further from the MCP-PMTs, and for photon paths that undergo reflections from the side and/or bottom of the radiator, 
a dependence on the path-length of the photons is expected, however the dependence seen in data is worse than predicted by simulation. 
This performance should be improved by future calibrations of the MCP-PMT readout electronics. 
The measurement of the time resolution of the charged track as a function of number of photon hits needs further study, and to understand the effect of clustering and correlations of closely-spaced hits of photons within the MCP-PMTs on the expected performance.

A test beam campaign is planned for the end of 2022, which will employ the fully instrumented TORCH prototype (with up to eleven MCP-PMTs). This will be used to further verify the photon yield and  time resolution performance, and a mixed charged-particle beam  will be used to measure the time-of-flight PID separation of different particle species over a range of momenta. 

\section*{Acknowledgements}
The support is acknowledged of the Science and Technology Research Council, UK, grant number ST/P002692/1, and of the European Research Council through an FP7 Advanced Grant (ERC-2011-AdG299175-TORCH).
We acknowledge the contributions from the CERN T9 beam operators. 
We would like to thank Ankush Mitra, Jeanette Chattaway of the University of Warwick, Keith Clark of the University of Bristol,    Mark Jones and Peter Hastings at the University of Oxford,  and the staff of the  workshops at the Universities of Bristol, Edinburgh, Oxford and Warwick Departments of Physics for their help in the production of the Proto-TORCH mechanical frame and electronics. 
We also thank Jon Lapington of the University of Leicester for his advice and insightful discussions on the MCP-PMTs and Chris Slatter of Photek for his work on their development. An individual author (TB) has received support from the Royal Society (UK).


\bibliographystyle{LHCb} 
\bibliography{refs}

\end{document}